\documentclass[10pt,journal,compsoc]{IEEEtran}
\usepackage{color}
\usepackage{graphicx}
\usepackage{tabularx}
\usepackage{hyperref}
\usepackage{algorithm,algorithmic}
\usepackage{comment}
\usepackage{booktabs}
\usepackage{enumitem}
\usepackage{amssymb}
\usepackage[normalem]{ulem}

\newcommand{\add}[1]   {\textcolor{red}{#1}}

\usepackage{xcolor}
\usepackage{soul}
\usepackage{booktabs}
\usepackage{comment}
\usepackage{multirow}

\newcounter{example}[section]
\newenvironment{example}[1][]{\refstepcounter{example}\par\medskip
   \textbf{\textit{Example~\theexample. #1}} \rmfamily}{\medskip}

\newcolumntype{L}{>{\raggedleft\arraybackslash}X}
\newcolumntype{R}{>{\raggedright\arraybackslash}X}

\ifCLASSOPTIONcompsoc
  \usepackage[nocompress]{cite}
\else
  \usepackage{cite}
\fi
\ifCLASSINFOpdf
\else
\fi
\begin{document}
\title{Improved management of issue dependencies in issue trackers of large collaborative projects}

\author{Mikko~Raatikainen, %~\IEEEmembership{Member,~IEEE,}
        Quim~Motger, %~\IEEEmembership{Fellow,~OSA,}
       Clara~Marie~L\"uders,
        Xavier~Franch,
        Lalli~Myllyaho,
        Elina~Kettunen,
        Jordi~Marco,
        Juha~Tiihonen,
        Mikko~Halonen,
        and~Tomi~M\"annist\"o % <-this % stops a space
\IEEEcompsocitemizethanks{\IEEEcompsocthanksitem M. Raatikainen, L. Myllyaho, E. Kettunen, J. Tiihonen, and T. M\"annist\"o are with University of Helsinki, Finland.
 E-mail: mikko.raatikainen@helsinki.fi, lalli.myllyaho@helsinki.fi, elina.kettunen@alumni.helsinki.fi, juha.tiihonen@alumni.helsinki.fi tomi.mannisto@helsinki.fi.
\IEEEcompsocthanksitem Q. Motger, X. Franch, and J. Marco are with Universitat Politècnica de Catalunya, Barcelona, Spain.  E-mail: jmotger@essi.upc.edu, franch@essi.upc.edu, jmarco@cs.upc.edu.
\IEEEcompsocthanksitem  C. M. L\"uders is with University of Hamburg, Germany. E-mail: lueders@informatik.uni-hamburg.de.  
\IEEEcompsocthanksitem M. Halonen is with The Qt Company, Oulu, Finland. E-mail: mikko.halonen@qt.io.}
\thanks{Manuscript submitted March, 2022.}
\thanks{Corresponding author: M. Raatikainen}}


\markboth{Manuscript accepted to IEEE TRANSACTIONS ON SOFTWARE ENGINEERING:  doi: 10.1109/TSE.2022.321216}%
{Manuscript accepted to IEEE TRANSACTIONS ON SOFTWARE ENGINEERING:  doi: 10.1109/TSE.2022.321216}



\IEEEtitleabstractindextext{
\begin{abstract}
Issue trackers, such as Jira, have become the prevalent collaborative tools in software engineering for managing issues, such as requirements, development tasks, and software bugs.
However, issue trackers inherently focus on the lifecycle of single issues, although issues have and express dependencies on other issues that constitute issue dependency networks in large complex collaborative projects. 
The objective of this study is to develop supportive solutions for the improved management of dependent issues in an issue tracker.
This study follows the Design Science methodology, consisting of eliciting drawbacks and constructing and evaluating a solution and system. The study was carried out in the context of The Qt Company's Jira, which exemplifies an actively used, almost two-decade-old issue tracker with over 100,000 issues. 
The drawbacks capture how users operate with issue trackers to handle issue information in large, collaborative, and long-lived projects. The basis of the solution is to keep issues and dependencies as separate objects and automatically construct an issue graph. Dependency detections complement the issue graph by proposing missing dependencies, while consistency checks and diagnoses identify conflicting issue priorities and release assignments. Jira’s plugin and service-based system architecture realize the functional and quality concerns of the system implementation. 
We show how to adopt the intelligent supporting techniques of an issue tracker in a complex use context and a large data-set. The solution considers an integrated and holistic system view, practical applicability and utility, and the practical characteristics of issue data, such as inherent incompleteness.
\end{abstract}

}

\maketitle

\IEEEdisplaynontitleabstractindextext

% \fi
\IEEEpeerreviewmaketitle

\IEEEraisesectionheading{\section{Introduction and motivation}\label{sec:introduction}}

\emph{Issue management} is a fundamental activity in many of today's software development projects and is especially prevalent in open source projects. It consists of identifying and resolving new requirements, development tasks, unexpected problems, software bugs, and questions (i.e., the \emph{issues}) that may arise during the project. As issues convey important observations, failing to manage issues may result in delays, quality problems, or even complete software project failure~\cite{Bertram2010}.
Due to this critical nature and complexity, particularly in large collaborative projects, issue management is usually tool-supported: software engineering teams use \emph{issue trackers} to report, manage, and resolve software project-related issues~\cite{Bissyande2013}. Issue trackers are typically used collaboratively by various project stakeholders, including project managers, developers, and even end-users.

In this complex, collaborative environment, issues cannot be conceived as independent entities, although users report them individually. Instead, issues affect each other through various \emph{dependencies}, forming an \emph{issue dependency network}. For example, a reported issue might be part of a major bug, a bug might contribute to a specific requirement, or two issues might refer to the same topic.
Dependencies are a critical concern that various software engineering planning activities, such as requirements prioritization~\cite{Achimugu2014}~\cite{Thakurta2016} 
and release planning~\cite{Svahnberg2010}~\cite{Ameller2016}, need to consider. In this paper, we refer to dependencies specifically as horizontal interdependencies between issues, rather than vertical dependencies, i.e., traceability between issues and other types of artifacts, such as issues and their implementation~\cite{Dahlstedt2005}.

The main challenges arise from how issue trackers create and manage the dependencies between the issues. Issue trackers record individual issues to provide stakeholders with information over the issue's life cycle, and the dependencies are always bound to a specific issue.
Therefore, overall understanding of and advanced analytical and management features for issue dependency networks are not well supported. Moreover, users do not always report dependencies properly because it is manual and tedious. This lack of dependencies results in the possibility of developers resolving an issue without being aware of its dependencies.
Given that issues rarely appear in isolation, these limitations in managing issue dependency networks are harmful to development and product management.

In this paper, we elaborate on the challenges and the construction and application of solutions to support software project stakeholders. This support focuses on managing dependent issues in an issue tracker over the development life cycle. The research follows the Design Science methodology~\cite{Gregor2006,Hevner2004}, addressing the problem, solution knowledge, and artifact instance realization as applied research. Our solution focuses on dependency management baseline techniques by formalizing dependencies and techniques for detecting missing dependencies and consistency analysis of issue dependency networks to extend well-known features offered by issue trackers.
The implementation of the solution has a Jira plugin as its user interface, and independent microservices realize issue dependency management techniques. The solution design considers contextual product quality characteristics -- including security, scalability, and efficiency -- to fit in real, large data-set scenarios. To this end, the problem analysis, solution design, and evaluation have been carried out iteratively in the context of \emph{The Qt Company (TQC)}, a publicly listed global software company. However, we examine the results with the purpose of generalizing our contributions beyond the TQC case.

This paper is organized as follows. 
Section~\ref{section:background} provides background about issue trackers. Section~\ref{section:researchMethod} describes the overall research method, including the research questions and a description of TQC and its Jira. The results are presented in three sections: Section~\ref{section:currentDrawbacks} reports the main drawbacks in issue tracker use, Section~\ref{section:drawbacksManagement} depicts the objectives and techniques of our solution, and Section~\ref{section:implementation} addresses the artifact implementation. Section~\ref{section:quality} reports the evaluation. Section~\ref{section:discussion} collects discussion and related work. Section~\ref{sec:Validity} discusses the threats to validity. Finally, Section~\ref{section:conclusions} concludes the research.

\section{Background: Issue trackers}
\label{section:background}

Issue trackers provide technological support for issue management tasks as they document and communicate knowledge. To this end, most issue trackers have additional features that allow users to comment, watch, or like specific issues. Given the collaborative nature of these tools and their features, they are \textit{"a type of social media"}~\cite{Chrupala2012} for the software development domain.

Karre et al.~\cite{Karre2017} analyzed 31 well-known issue trackers to identify their features and main differences. The analysis resulted in 24 characteristics, including simple, traditional features, such as e-mail support or the existence of a comments section, and more complex features, such as a customizable graphical user interface or the ability to establish links between independent issues. 
Among the most advanced issue trackers, which contain complex features like issue field customization, release planning, and project management features, we highlight Jira, Redmine, Mantis, Bugzilla,  and GitHub\footnote{Based on the latest GitHub issues release: https://github.com/features/issues}. For these, we have built a comparative analysis based on a high-level functionality coverage (Table \ref{tab:its-comparative-analysis}). These advanced issue trackers are all well-known and widely-used tools that provide sophisticated issue modeling features and include a wide variety of issue management functionalities, especially for single-issue management. These single-issue modeling features include type (e.g., `epic', `bug', `user story', and `task'), scope (product or component), and status (e.g., `open' and `closed'). However, there are significant differences among them. For instance, neither Bugzilla nor Mantis support the definition of custom issue types or statuses nor issue types other than `bugs', which is the underlying type of each issue. On the other hand, these are features supported by both Redmine and Jira.

If we focus on more advanced features beyond single-entity analysis, 
all advanced issue trackers support the specification process for dependencies among issues (i.e., issues depending on the resolution of another issue) or duplicated issues (i.e., marking an issue as a copy of an existing one). However, these dependency and duplicate management features are limited to the specifications of properties of a single issue. In addition, this complex specification process requires human action to label and create these dependencies manually.

Some advanced features, such as release management tasks, including creating a release plan and adding issues to a scheduled release, are supported only by Jira and, recently, by GitHub. These features make Jira one of the most advanced issue trackers in terms of the scope of its functionalities.
Using TQC as a context to acquire more depth on the qualitative side of the problems, we can produce a  generalized case study account and findings from Jira and TQC comparable to other contexts and issue trackers. A more detailed account of the existing features in Jira is in Sections~\ref{section:researchMethod} and~\ref{section:currentDrawbacks}

\setlength{\tabcolsep}{3pt}
\begin{table}[]
\centering
\caption{{Comparative analysis of issue trackers based on functionality coverage}}
\begin{tabular}{lccccc}
\toprule
 & {Jira} & {Redmine} & {Mantis} & {Bugzilla} & {GitHub} \\
\midrule
{Issue dependencies} & {\checkmark} & {\checkmark} & {\checkmark} & {\checkmark} & {\checkmark} \\
{Dependencies typology} & {\checkmark} & {\checkmark} & {\checkmark} & {(\checkmark)\textsuperscript{a}} & {-}\\
{Duplicated issues} & {\checkmark} & {\checkmark} & {\checkmark} & {\checkmark} & {\checkmark} \\
{Release management} & {\checkmark} &{-} & {-} & {-} & {\checkmark} \\
{Custom fields}  & {\checkmark} & {\checkmark} & {\checkmark} & {\checkmark} & {\checkmark}\\
{Custom dashboards}  & {\checkmark} & {\checkmark} & {-} & {\checkmark} & {\checkmark} \\
{Custom workflows}  & {\checkmark} & {-} & {\checkmark} & {\checkmark} & {-}\\
{Add-ons/plugins} & {\checkmark} & {\checkmark} & {\checkmark} & {\checkmark} & {-}\\
{API integration} & {\checkmark} & {\checkmark} & {\checkmark} & {\checkmark} & {\checkmark}\\
\bottomrule
\end{tabular}
\label{tab:its-comparative-analysis}
\raggedright a) Harder to customize and not for the overall management, but different types (\emph{depends/requires}, \emph{see also}, and \emph{duplicate}) exist.
\end{table}
\setlength{\tabcolsep}{6pt}

\section{Research approach: Design Science}
\label{section:researchMethod}

Our research follows the Design Science methodology~\cite{Gregor2006,Hevner2004}. The objectives are to understand a relevant problem and construct solution knowledge that governs the principles and use of an artifact. The solution knowledge is applied to create and use an artifact instance in a specific context because the artifact instance might not have existed earlier or has at least not been used for solving the problem. The resulting artifact instance needs to be assessed to operate appropriately by solving the problem and have utility in the context, thus providing validation for the truthlikeness and utility of the solution knowledge and artifact~\cite{Niiniluoto1993}. The artifact instance and context, which in our case is a single case of TQC, are a contextual basis for analytical generalization.
This section describes how we applied Design Science in our research process and the context of TQC, where we carried out the research using a qualitative, applied research approach.

\begin{figure*}[t]
	\centering
		\includegraphics[width=\textwidth]{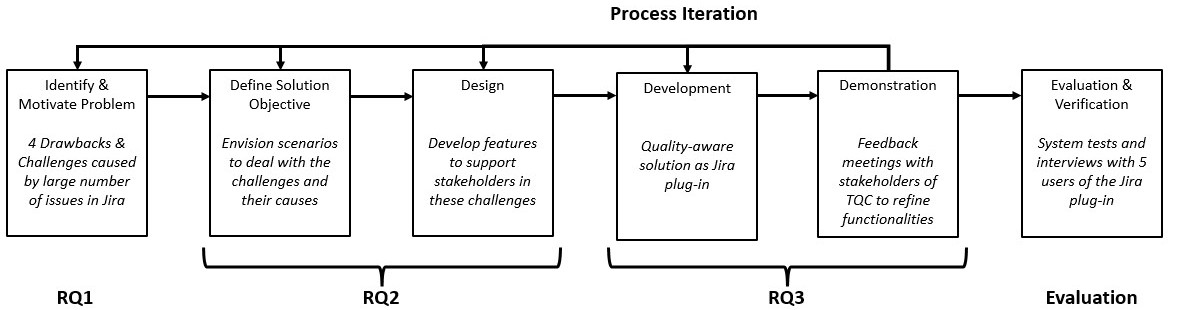}
	\caption{The phases of incremental and iterative Design Science process of Peffers \textit{et al.} ~\cite{Peffers2007} applied in this research.}
	\label{fig:researmethod}
\end{figure*}

\subsection{Research process}
We apply the Peffers \textit{et al.}~\cite{Peffers2007} incremental and iterative process to Design Science, as illustrated in Figure~\ref{fig:researmethod}. The phases are linked to the research questions  and evaluation, as shown at the  bottom of Figure~\ref{fig:researmethod}. As a part of process iterations illustrated on the top of Figure~\ref{fig:researmethod}, we frequently discussed the research with TQC's stakeholders and incorporated their feedback.
Specifically, we had an approximately bi-weekly online status meeting in which one of TQC's managers participated. The research questions (listed in Table~\ref{tab:rqs}) are:

\textbf{Problem identification (RQ1).}
We had conducted an exploratory, multiple case study to understand companies' pain points, needs, and challenges for a platform to support large-scale requirements engineering~\cite{Fucci2018} as part of the collaborative OpenReq research and innovation project\footnote{www.openreq.eu}. TQC was one of the five organizations in the study. Data were collected using semi-structured, half-day interviews with one product manager and two senior R\&D managers selected by a TQC representative due to their active involvement in maintaining and improving Jira and the RE process within TQC.
The participants worked in the Qt R\&D sites in Germany, where the interviews were conducted in English. Interviews were audio-recorded and later transcribed. The data were analyzed using open and axial coding on the Atlas.ti\footnote{www.atlasti.com} tool. One researcher held the primary responsibility, and other researchers supported and reviewed them.
The main problems found were: information overload, limited tool support, handling of dependencies between requirements, and stakeholder identification for issue assignment. The research method and results details are available at~\cite{Fucci2018}.

\begin{table}[t]
    \centering
    \caption{Research questions of the study}
    \label{tab:rqs}
    \begin{tabularx}{\columnwidth}{@{}lX@{}}
    \toprule
    ID & Text \\
    \midrule
    RQ1 & What drawbacks do stakeholders suffer with current issue trackers? \\[1ex]
    RQ2 & What features can be added to issue trackers to address these drawbacks? \\[1ex]
    RQ3 & How can these features be integrated in an issue tracker so that it has value for use? \\[1ex]
   % RQ 4 & What is the overall quality of the resulting artifact?\\
    \bottomrule
    \end{tabularx}
\end{table}

RQ1 of this paper refines the problems mentioned above in~\cite{Fucci2018} to better understand these problems rather than explore the broader problem space.
First, however, we excluded the problem of stakeholder identification because open source communities are generally sensitive to the disclosure of personal information. Then, the remaining three problems were analyzed in greater depth to gain a more holistic and detailed understanding and provide concrete examples. This analysis paralleled the process of the following RQs. We had access to the earlier interview data, met several times with two additional managers at TQC and another manager even more frequently who provided us with feedback and contextual information, and analyzed publicly available TQC Jira issues by ourselves and based on example issues mentioned by TQC's employees. This analysis of issues also triangulated and validated our understanding.  The existing triaging process, the ongoing requirement engineering process improvements, and a TQC manager's emphasis on challenges with dependencies, as detailed in Section~\ref{sec:ReseachContextTQC} below, made us focus on the functionality of Jira rather than process improvement.   
A respondent even pointed out challenges in Jira: "\textit{So I realize we’re sitting on huge knowledge bases, which is Jira, with lots and lots and lots of details, but often it feels like we are working on a haystack. So it’s just so overwhelming, that it’s hard sometimes to really, you know, find out what is important and what not. And having some help there would be great}" (sic)  later continuing "\textit{as I said, Jira is just a big haystack, and having some analytics to understand,... okay, the pain point}" (sic). 
The resulting drawbacks are summarized and provided with the same or similar challenging examples pointed out by TQC's employees in Section~\ref{section:currentDrawbacks}.

\textbf{Define solution objectives and design (RQ2).}
The solution objectives synthesize functionality to mitigate the drawbacks.  The design principles of solution techniques  realize the objectives and integrate with issue trackers as the tool and the existing issues of the issue trackers as the data.  As a central design principle, the solution should not change but instead support and complement current ways of working to lower the adoption barrier. 

While understanding the context and drawbacks (RQ1), we also synthesized the objectives of the solution. We held frequent meetings with the TQC managers to combine the problem domain understanding of TQC and our solution proposal knowledge. We then defined the solution techniques, including how different techniques were considered and even tested as prototype implementations. Some techniques were abandoned, e.g., because of challenges in practical scalability or required data quality. 
Overall, the process was iterative and selective, resulting in objectives and solution techniques in Section~\ref{section:drawbacksManagement}. 

\textbf{Implementation of the solution and demonstration of its operations (RQ3).}
The incremental development of the artifact to realize the solution was done iteratively with a feedback loop from TQC, which helped ensure that the artifact was meeting general quality objectives, which we structured following the ISO/IEC 25010 product quality model~\cite{ISO25010}. The implementation consists of a Jira plugin as a user interface and the solution techniques as independent microservices.
These challenges shaped the final artifact design, as detailed in Section~\ref{section:implementation}.

\textbf{Evaluation.}
\label{sec:ResearMehod-Evaluation}
The final evaluation is divided into \emph{verification} and \emph{validation}~\cite{ISO25010}. Verification evaluated the results against the stated objectives, and it was carried out by executing system tests that explored the functionality and observed and measured the product quality characteristics.  
Validation techniques assess the results with their intended users through interviews. 
The technical details and results of the evaluation are provided in Section~\ref{section:quality}.

\subsection{Research context: TQC and issue management}
\label{sec:ReseachContextTQC}
 
TQC, as the company is now known, was established in the 1990s, later acquired by Nokia Plc and Digia Plc, and now operates as an independent publicly listed company. TQC is growing fast with around 300 employees and major R\&D sites in Finland, Germany, and Norway. TQC is a global software product and service company applying typical modern software engineering tools and methods, such as agile practices.

TQC governs the Qt project (hereafter \emph{Qt}), which is developing 
a software development kit\footnote{www.qt.io} consisting of the Qt software framework itself and its supporting tools. These tools include the integrated development environment~(IDE) called Creator and the 3D Studio (3DS) and extensions to the Qt software framework, such as the Automotive suite. Qt specifically targets the development of cross-platform mobile applications, graphical user interfaces, and embedded applications. Qt is estimated to be used by one million developers and most of today's embedded and touch screen systems rely on Qt. Qt is available under open source and commercial licenses. 

Jira is the only system for product management and requirements engineering. The oldest issue in Jira dates back to the 23rd of September 2003, i.e., almost two decades old. Each Jira issue has an ID consisting of a preceding project acronym and a running number (e.g., `QBS-991'\footnote{https://bugreports.qt.io/browse/QBS-991}), a title (`Qt Android support') and description, as well as several properties, such as the type (in QBS-991, a bug), release (referred to as \textit{Fix Version/s}), priority, status (identifies where an issue is in its lifecycle, such as `Open', `Closed'), resolution (gives additional details for status, such as an issue is closed because it is a `Duplicate'), and automatic meta-data, such as the creation date. There are various releases, such as major and minor releases and bug fixes, and the release numbering typically follows up to three-part (x.y.z). Priority ranges from 0 (`P0 blocker') to 5 (`P5 not important'). In addition, an issue includes comments.

In TQC's Jira, issues may report `bugs', `epics', `user stories', `suggestions', and `tasks'. While bugs are the prevalent issues, TQC aims to organize development by applying an issue hierarchy like in agile methods: large functionalities or features are defined as `epics' that are refined as User Stories and further as `tasks'. TQC underwent ongoing requirement engineering process improvement that aimed to enforce this hierarchy further. In addition to the parent-child relationships induced by this issue hierarchy, issues can have dependencies referred to as \textit{Issue Links} in Jira. These links can only be set by employees of TQC or authorized open source developers. Other TQC Jira users, even the creators of issues, cannot set any links. TQC's Jira supports the following links: `duplicates', `requires', `relates', `replaces', `results', and `tests'. All these links are bidirectional (e.g., `is related to' and `relates to'), but it is not uncommon for users to declare a wrong direction, especially in the case of a duplicate, as the resolution already shows duplication.  There are also several exceptions or misuses for these types. Sometimes issues are used only to gather other issues, such as one major epic depending on other epics, as epics cannot form a parent-child hierarchy (e.g., QTBUG-62425\footnote{https://bugreports.qt.io/browse/QTBUG-62425}).
Some issues group other issues in the description or comments field (e.g., QTCOMPONENTS-200\footnote{https://bugreports.qt.io/browse/QTCOMPONENTS-200}), and not all of them are linked in the appropriate fields.
Although Jira handles parent-child relationships and links differently, we simply use \textit{dependency} to denote both.

TQC's Jira is divided into projects. Examples include: `QTBUG', which contains issues related to the Qt Framework, and `QTCREATORBUG', which contains issues related to Creator. The large projects are further divided into components, such as a Bluetooth component in `QTBUG'. Each component has a responsible maintainer from TQC's R\&D department or the open source community. TQC's product management has more general responsibility for the projects. The projects and components have dependencies, including cross-project dependencies, like the Automotive suite built on top of Qt Framework.

\begin{table}											
\caption{The number of issues and dependencies in the three largest and other projects in total on the 29th November 2019.}
\label{tab:JiraData}		%\centering							
\begin{tabularx}{\columnwidth}{X 
>{\raggedleft\arraybackslash}p{1.71cm} 
>{\raggedleft\arraybackslash}p{1.71cm}
>{\raggedleft\arraybackslash}p{1.71cm}
% >{\raggedleft\arraybackslash}p{1.25cm}
}									
% % \begin{tabularx}{\columnwidth}{p{2cm} >{\raggedleft\arraybackslash}p{.84cm} 
% % >{\raggedleft\arraybackslash}p{1.2cm}
% % >{\raggedleft\arraybackslash}X >{\raggedleft\arraybackslash}p{1.4cm}}										
% 	&	\textbf{Qt framework}	&	\textbf{Creator}	&	\textbf{3DS}	&	\textbf{Other 17 projects}	\\
% \toprule			
% & & & &  \\[-.25cm]
% Issues	&	78,676	&	21,926	&	3,877	&	15,441	\\
% Internal \newline dependency	&	15,739	&	3,126	&	2,023	&	3,517	\\
% Cross-project dependency	&	1,811	&	1,132	&	133	&	1,307	\\

	&	\textbf{Issues}	&	\textbf{Internal \newline dependencies}	&	\textbf{Cross-project \newline dependencies*}	\\
\toprule			
& & &  \\[-.3cm]
Qt Framework	&	78,676	&	15,739	&	1,811 \\
Creator &	21,926	&	3,126	&	1,132 \\
3D Studio	&	3,877	&	2,023	&	133 \\
Other projects	&	15,441	&	3,517	&	1,307 \\
\bottomrule	
\end{tabularx}
* A dependency between two projects is counted in both projects.
							\end{table}

Anyone can register and report issues to TQC's Jira and view the full details of issues, follow issues, and add comments. Qt operates in a meritocratic manner, in which developers get promoted as Jira users when they contribute to Qt and receive recommendations from other developers. Only those who have received elevated rights can edit issues. This preserves quality and  integrity.

All new issues go through a triaging process\footnote{https://wiki.qt.io/Triaging\_Bugs} to ensure issue quality. Triaging includes checking the quality of the issue in terms of relevance and understandability, specifying which component the issue concerns, assigning the responsible maintainer, and setting the priority. 
Typically, issues are triaged within a few days at most. The triaging responsibility is rotated within TQC, but the maintainers or others with proper elevated rights can also triage and manage issues before TQC's responsible employees react.

To monitor overall progress, TQC uses dashboards for each release with swim lanes for status categories `not started', `in progress', `blocked', and `done'. A dashboard is a feature in Jira to automatically filter, organize, and visualize a set of Jira issues based on their property values, such as the above release and status.

TQC's Jira is an independent deployment in a virtual machine in the Amazon cloud services. In addition, TQC has a snapshot of this virtual machine as a test environment, which we use in our research. The snapshot was taken on the 29th November 2019.  In this snapshot, TQC's Jira is divided into~20 public, separate projects. We used this same data snapshot for all tests to make the results comparable.
Table~\ref{tab:JiraData} shows the number of issues and dependencies in Qt Framework, Creator, and 3D Studio, the three largest projects, and the remaining 17 other projects combined. Out of a total of 119,920 issues, 26,746~(22\%) issues were modified within the last year before the test snapshot (29.11.2018-29.11.2019), and~25,938~(22\%) were open, i.e., not resolved, at the end of the period. Modifications include any changes, such as editing text, changing properties, or adding comments.
In addition, TQC has about ten private projects in Jira for specific customers and product management, which contain a few thousand additional issues. For confidentiality reasons, these projects are not included in the data-set of this paper.

\section{RQ1: Drawbacks in issue management}
\label{section:currentDrawbacks}

RQ1 describes  drawbacks related to using  issue trackers' functionalities as they appear in Jira and at TQC.

\textbf{\emph{Drawback 1.} Limited view of the issue dependency network.} 
When resolving an issue, issue tracker users typically need to consider the issue dependency network. Considering both the parent-child relation of \textit{epic -- user story -- task} hierarchy and issue links, Jira issue dependencies constitute a set of large, disconnected issue networks. To explore a resulting issue dependency network beyond direct dependencies, a user needs to follow the dependencies from one issue to another. The drawback is that it is tedious and error-prone for users to form an overall understanding of any larger network structure by following the links one by one because Jira -- or other advanced issue trackers --  does not support other ways to explore an issue dependency network. Moreover, none of the other advanced issue tracker functionalities, such as searches and dashboards in the case of Jira, can take dependencies into account automatically.

\begin{example}
Issue QT3DS-1802 has 15 dependencies to other issues, which, in turn, have 59 additional direct dependencies. The network grows further similarly beyond these issues. A Jira user needs to open each dependent issue to see its details and how many (if any) dependencies there are in these dependent issues beyond direct dependencies. In the worst case, this means opening dozens of issues separately and keeping in mind what is dependent on what and how because there is no single view over the dependency network.
The largest issue network consists of 8,952 issues. There are 270 networks with more than seven issues contained in them.
 \end{example}

\textbf{\emph{Drawback 2.} Issues lack explicit dependencies.}
Issue trackers require users to report dependencies among issues manually. Eventually, users may not report all of them, resulting in missing dependencies, which may have critical consequences for activities like ensuring the integrity and quality of a release. Jira users have reported that this is a frequent situation and identified five different reasons behind missing dependencies:
\begin{itemize}
    \item {\it Unawareness.} When reporting an issue, a user is unaware of all related issues. 
    \item {\it Uncertainty.} A user is unsure whether a specific dependency is needed or not. It is customary that uncertain dependencies are only mentioned in the description or comments of an issue rather than appropriately reported.
    \item {\it Discrepancy.} Users have different opinions on whether or not an explicit dependency is needed. 
    \item {\it Lack of time.} Even when a user is entirely sure about a dependency, adding it can be cumbersome, requiring several steps. 
    \item {\it Lack of permissions.} Not everyone is authorized to add dependencies, rather the operation requires elevated privileges.
\end{itemize}

\begin{example}
\label{ex:reference}
A TQC Jira user had added a comment on the issue QBS-881: ”i see this task as being redundant with QBS-912 - close?”
(sic). Another user  agreed in a follow-up
comment. However, no one reported an explicit dependency.
Such comments often go unnoticed and users looking at QBS-912 are unaware of the comments in QBS-881. To understand the magnitude of possibly missing or incorrectly reported dependencies, we emphasize the Qt 3D Studio project as an example.  A developer of Qt 3D Studio stated that they aim to use dependencies rigorously. As a result, 50\% of the issues in the project have dependencies compared to 25\% in Qt Framework and 24\% in Qt Creator.  
\end{example}

\textbf{\emph{Drawback 3.} Duplicated issues.} As anyone can create an issue, it is not uncommon that the same topic is reported more than once, resulting in very similar issues, which can be considered duplicates.  As already found in~\cite{Fucci2018}, issue tracker users reported that it would be convenient to detect and link duplicates to comprehend the issue network structure better.
On the other hand, it is also important not to delete any of them because each issue can still have some original content. 

Jira offers `duplicate' as a resolution property value to indicate that the issue duplicates another issue and the `duplicates' dependency to connect duplicate issues. Any issue that duplicates another issue should have a `duplicates' dependency towards the duplicated issue and have the resolution and status properties marked as `duplicate' and `done', respectively. 
However, these tasks and initially identifying the duplicates need to be done manually and thus are often neglected.

Since duplicate dependencies are a type of dependency, the reasons for and consequences of missing duplicates are similar to the previous drawback. Another shortcoming is that the community can voice their opinion on issues by watching them, which indicates the popularity of an issue. If there are duplicates of an issue and the watchers are split over all of them, no one will be able to hear the community voice properly since that voice is incoherent. Thus, important information goes missing unless duplicates are detected. 

\begin{example}
A missing duplicate dependency is already exemplified in Example~\ref{ex:reference}. To illustrate challenges in the prevalence of duplicates and their management, TQC's Jira already has 8,150 (7\%) issues marked as `duplicate', of which 5,839 lack a `duplicates' dependency. 4,925 of these issues have some other dependency, which in some cases can mean that, e.g., a `relates' dependency is used incorrectly to denote duplication. Still, the remaining 914 issues do not have any dependency. In addition, it is possible that duplicate issues have simply been closed without setting the resolution or that some duplicates have gone unnoticed.

\end{example} 

\textbf{\emph{Drawback 4}. Incorrect release assignments and priorities in an issue dependency network.}
As software systems have specific release cycles, it is relevant to consider the issue dependency network when deciding on resolving issues.
For example, \emph{A} \textit{requires} \emph{B} dependency means that the solution of \emph{A} needs the solution of \emph{B} to operate properly. Thus, it is not reasonable to release \emph{A} first as its solution will not be useful without \emph{B}. Advanced issue trackers offer progress monitoring features for release plans, such as dashboards and filters. However, the features do not currently  offer any automatic checking of the logical correctness of assigned releases over dependencies, and all checks need to be done manually.  Any fault that goes unnoticed can lead to an incomplete release.

\begin{example}
TQC's Jira users reported two practically relevant \textit{dependency rules} for their context:
\begin{itemize}
    \item \emph{Parent-child rule}. In a parent-child dependency, the child must be scheduled in the same or an earlier release than its parent or have a lower priority.
    \item \emph{Requires rule}. A required issue must not have a later release or lower priority than an issue requiring it. 
\end{itemize}

An example of an incorrect release version with a 'requires' rule is the issue ``QTBUG-72510''. It has the release version 5.13 and a sub-task that is not assigned to any release.  An example of an incorrect priority is``QTBUG-27426'' (with priority P0), requiring ``QTBUG-28416'' (priority P2), which violates the rule that a required issue cannot have a lower priority. 
Over 12\% of all `requires' and `parent-child' inter-project dependencies in TQC's Jira violate the dependency rules.
\end{example}

\section{RQ2: Objectives and Features for the enrichment of issue management}
\label{section:drawbacksManagement}
In this section, we cover the solution objectives and then the baseline and concrete techniques of our solutions.

\subsection{Objectives}
\label{sec:objectives}

Based on the drawbacks enumerated in the previous section, we present the synthesized solution objectives, which aim to alleviate the drawbacks and improve dependency management in issue trackers.
\begin{itemize}
    \item \textit{\textbf{Objective 1.}} Users gain a better understanding of the existing issue dependency network of the issues (\textit{Drawback 1}).
    \item \textit{\textbf{Objective 2.}} Users can search for missing dependencies and unidentified duplicate issues (\textit{Drawback 2}, \textit{Drawback 3}).
    \item \textit{\textbf{Objective 3.}} Users can check the correct release assignments and priorities of the issue dependency network of issues and receive suggestions for resolving inconsistencies (\textit{Drawback 4}). 
\end{itemize}

These objectives share four common, additional characteristics. First, the objectives integrate into the current ways of working with issue trackers, being usable whenever needed without disturbing existing processes.  
Second, the objectives are about improving issue trackers so that their realization becomes integrated into the functionalities and especially data of issue trackers.  
Third, the objectives address the context of the existing issues that the user is working on.
Fourth, the objectives should be realizable through scalable solutions, capable of working efficiently even with large projects.
Consequently, the objectives primarily address tool improvement rather than process improvements or changes.

\subsection{Dependency management baseline techniques}
\label{sec:drawbacksManagement-background}

The fundamental principle of our solution is that the roles of dependencies in issue trackers can be first-class entities rather than only properties of issues. We approached this by handling issues and dependencies as two distinct entity types in a graph-like structure: issues are nodes, and dependencies are typed (i.e., labeled) and directed edges between the nodes. This approach gives issues a context beyond their explicit properties, revealing implicit constraints, e.g., the mutual aggregation of two issues through a dependency between them. Moreover, dependencies can then have properties of their own, like issues have, such as a status and creation date. 

We define what we call an \emph{issue graph} as follows.
We denote the set of all issues as $R$ and the set of all dependencies between issues of $R$ as $D$, i.e., ${D \subseteq R \times R}$, where $D$ is anti-reflexive, i.e., ${\forall r_i \in R : (r_i,r_i) \notin D}$; and all edges are bidirected, i.e. ${\forall r_i, r_j \in R : (r_i,r_j) \in D} \iff (r_j,r_i) \in D$. That is, for every edge that belongs to the graph, there is also a corresponding inverse edge where the semantics of the edge depends on the direction.
For a particular issue $r_0 \in R$, the issue graph is a symmetric connected graph ${G_0 = (R_0, D_0)}$, where $R_0 \subseteq R$ and $D_0 \subseteq D$,
so that all issues of $R_0$ are reachable from $r_0$, i.e., for all issues $r_i \in R_0$ there is a path from $r_0$ to $r_i$ %, denoted as $p(r_0, r_i) = (r_0,...,r_i)$. %, 
and $D_0$ includes all dependencies between the issues in $R_0$ and only those. 
This definition of an issue graph is issue-centered and does not necessarily include all issues ($R_0 \subsetneq R$) because there is normally no path between all issues. However, the union of all $G_0$, denoted by $G = \bigcup G_0$, contains all issues ($R$) and dependencies ($D$). Equivalently, every $G_0$ is a component of $G$.
A special case of $G_0$ is an \textit{orphan issue} $r_0$ with no dependencies, and thus for which $R_0 = {r_0}$ and $D_0 = \emptyset$. 

Given an issue $r_0$, we define $G^p_0$, called a \textit{$p$-depth issue graph}, as an induced subgraph of $G_0$ that includes all issues up to $p$ edges apart from $r_0$ and all dependencies between the included issues.
That is, an issue is taken to the point of focus, and we follow all dependencies of that issue to neighboring issues and beyond, breadth-first up to the desired depth. 
The rationale and benefit of a $p$-depth issue graph are that different sizes of \textit{contexts of analysis} can be constructed automatically without user involvement to provide a given issue with the issues and dependencies in specific proximity. 

For an issue~$r_i$, we can apply a number of functions, such as $r_i.property(priority)$, to obtain its priority and $r_i.property(release)$  to get its scheduled release. Similarly, $d_i.property(status)$ will yield the status-property of a dependency~$d_i$, with possible values `proposed', `accepted' or `rejected' and $d_i.property(score)$ will give a score value (0..1) representing the confidence level of correctness or validity of the dependency.

These definitions provide the baseline for formulating the dependency management techniques required by users of issue trackers, addressing the objectives presented in Section~\ref{sec:objectives}. An issue graph ($G_0$) -- or the issue graph corresponding to the entire issue dependency network ($G$) -- can be generated automatically with the information stored in issue trackers; therefore, any operation defined over an issue graph or any transformation to any other formalism (e.g., constraint satisfaction problem (CSP)) can be computed from issue trackers, as we have effectively done. An issue graph does not necessarily  need to affect issue trackers; instead, the graph can form a parallel, complementary structure.

In particular, an issue graph ($G_0$)  makes efficient issue management  easier. Specifically, visualizations of dependency networks (\textit{Objective1}) can be generated directly for the user interface as described in Section~\ref{section:implementation}, and advanced dependency management techniques can be designed as described below.

\subsection{Dependency Management Techniques}
\label{sec:drawbacksManagement-techniques}

This subsection describes four concrete techniques of our solution, relying on the baseline techniques built on the concept of an issue graph. These techniques have been designed with the objectives in Section~\ref{sec:objectives} in mind. In particular, the techniques need to work in an industrial context, as exemplified in this paper by the TQC context, such as providing near real-time response times even when managing large sets of issues, which may sometimes prevent the adoption of more sophisticated approaches.

\textbf{Automated detection of potential missing dependencies.} 
As issue tracker users may neglect to report a significant number of dependencies, users would benefit from automatic dependency detection (\textit{Objective2}). Automation mitigates the burden of searching for the dependent issues, making it less critical for users to be familiar with all other existing issues. It is possible to automatically detect missing dependencies using various techniques, including deep learning~\cite{Guo2017}, active learning, and ontology-based approaches~\cite{Deshpande2020}.

In order to select an adequate dependency detection technique, we built on two fundamental observations. First, we prototyped a few complex techniques and observed that they did not meet the stringent time requirements and lacked proper training data. Second, TQC’s Jira users noted that dependencies are often mentioned as a reference to another issue in the title, description, and comments of an issue (Section~\ref{section:currentDrawbacks}, see Example~\ref{ex:reference}). Therefore, we adopted a \textit{reference detection} technique for natural language text (see Algorithm~\ref{algorithm:cross-det}) to uncover dependencies declared in the issue tracker in any of these text fields. In other words, our solution does not aim to identify unknown dependencies.

The reference detection technique analyzes this textually added content by searching for sub-strings consisting of a project acronym, a dash, and an integer (e.g., ''QBS-991'', see Section~\ref{sec:ReseachContextTQC}) representing an issue ID (line~4 of Algorithm~\ref{algorithm:cross-det}), and creates proposals for new dependencies whenever other issues are mentioned (lines~5--7). 
The reference detection technique marks the dependencies found as `proposed' (line~6) without type because only references to other issues are detected. 
A user should check, i.e., accept or reject these results, add the correct dependency type, and edit the textual content of the issue in case the proposed dependency is incorrect.

\begin{algorithm}[t]
 \caption{ReferenceDetection($R$, $projectID$)}
% \textit{g = (R, D)}: a directed issue graph \\
 \textit{R}: Set of issues of an issue graph \\
 \textit{projectID}: Set of project IDs (e.g., "QTWB", "QTBUG")
 
 \begin{algorithmic}[1]
  \STATE \textit{$D_{p}$} = []: set of proposed dependencies
  \FORALL{$r_i$ in $R$}
  \FORALL{$p_i$ in $projectID$}
  \STATE $toID[]$ = $r_i$.findStrings($p_i$+``-''+[0-9]\{1,5\})
  \FORALL{$to_i$ in $toID$}
  %\STATE $d = (r_i, r_j,`duplicates', `proposed')$
  \STATE $D_{p}$.add($r_i$, $to_i$, 'dependency', `proposed')
%   \STATE d=new($r_i$.ID, $to_i$)
%   \STATE d.type(`dependency').status(`proposed'))
%   \STATE $D_{p}$.add(d)
  \ENDFOR
  \ENDFOR
  \ENDFOR
  \RETURN $D_{p}$
 \end{algorithmic}
 \label{algorithm:cross-det}
\end{algorithm}

\textbf{Automated detection of potential duplicated issues.} The need for, the solution to, and the benefits of automatic duplication detection are much like the above because, as already noted in Section~\ref{section:currentDrawbacks}, duplicates result in a particular type of dependency in Jira (\textit{Objective2}). 
State-of-the-practice approaches use bag-of-words of natural language representations to measure the similarity between these representations using vector-space models~\cite{Shahmirzadi2019TextSI}. Among these approaches, \textit{Term Frequency - Inverse Document Frequency} (TF-IDF) is the theoretical baseline for  detecting duplicated entities or issues~\cite{Sun2011,Motger2020}. More recent deep contextualized models, such as Google's BERT~\cite{Devlin2019} or ELMo~\cite{Peters2018}, are more suitable for complex information retrieval scenarios, but as in dependency detection, they introduce a challenge in terms of efficiency, complexity, and training data required~\cite{Wang2019}. These challenges make it difficult to use them in the context of large issue dependency networks, as TQC exemplifies. 

Our solution (see Algorithm~\ref{algorithm:sim-det}) is an application and extension of the TF-IDF model based on three additional steps to improve the accuracy and performance of the similarity evaluation. After initially running the title and description of each issue through a lexical analysis pipeline (Lines 1--4 of Algorithm~\ref{algorithm:sim-det}), we built a TF-IDF model from the resulting bag-of-words representations (Line 5). Then, we apply the cosine similarity for the resulting TF-IDF model to compare each pair of issues. 
Alternative measures (e.g., Jaccard similarity) were evaluated through a preliminary literature review of the similarity evaluation field~\cite{Furnari2018}, which conveyed that TF-IDF and cosine similarity were suitable candidates for industrial scalability and generalization beyond the TQC case study.
Each resulting score is then compared to a context-based minimum threshold value to decide whether a pair is a potential duplicate, in which case a new `duplicate' dependency proposal is constructed (Lines 6--11). This context-based similarity threshold score must be set through a preliminary cross-validation analysis on a subset of labeled issues to predict duplicated issues accurately (see Section \ref{sec:DependencyManagementVerification}).

After the similarity evaluation, we represent the duplicated issues as sets of complete graphs, where issues have an existing or proposed `duplicate' dependency to other issues. We treat these sets of graphs as \textit{clusters} --- the process proposes sets of duplicated issues by simply including the issues belonging to the same cluster (Lines 12-13). During this process, we apply transitivity through existing duplicate dependencies to all issues belonging to the same cluster, resulting in new duplicated proposals. Hence, instead of reporting all the existing and proposed `duplicate' dependencies among them, we only report the duplicated dependency with the greatest similarity score for all other issues in the cluster. Given a sub-graph of $m$ duplicated issues, the clusters can be reported using $(m - 1)$ dependencies instead of representing all ($m*(m - 1)/2$) dependency objects, improving performance efficiency in data processing and transactions. The extension of the similarity algorithm through a clustering representation, the need for which was identified during construction,  improves the current state-of-the-art approaches by significantly reducing the amount of noisy data and simplifying the formalization of duplicated issues, similarly to up-to-date alternative solutions in the field \cite{Rocha2021}. Furthermore, stakeholders do not want all duplicate links explicitly in their issue trackers. A detection model might recommend all possible duplicate links, which the stakeholders do not want.

\begin{algorithm}[t] 
 \caption{DuplicateDetection($G$, $thr$)}
 \textit{G = (R, D)}: Issue graph \\
 \textit{thr:} Similarity threshold score
 \begin{algorithmic}[1]
  \STATE \textit{bow} = [] : Bag of words
  \STATE \textit{clusters} = [] : Set of sub-graphs of duplicated issues
  \FORALL{$r_i$ in R}
   \STATE \textit{bow}.add.text\_preprocess(\textit{$r_i$})
  \ENDFOR
  \STATE \textit{tfidf\_model} = build\_model(\textit{bow})
  \FORALL{$r_i$, $r_j$ $\in$ R where \textit{i $\neq$ j} and ${d_{ij} = (r_i, r_j)} \notin D$}
   \STATE \textit{score} = cosine\_sim(\textit{$r_i$, $r_j$, tfidf\_model})
   \IF{\textit{score} $\geq$ \textit{thr}}
  \STATE $D$.add($r_i$, $to_i$, 'duplicates', `proposed', \textit{score})
  \ENDIF
  \ENDFOR
  \STATE \textit{clusters} = compute\_clusters(\textit{R}, \textit{$D$})
  \RETURN clusters
 \end{algorithmic}
 \label{algorithm:sim-det}
\end{algorithm}

\begin{algorithm}[t]
 \caption{Proposals($r_0$, $D_0$, $D_0'$, $depth$, $orphan$, $property$)}
 \textit{$r_0$}: Issue of interest \\ 
 \textit{$D_0 = [(r_0,r_{p1}),...]$}: Dependencies for $r_0$ in TQC's Jira \\ 
 \textit{$D_0' = [(r_0,r'_{p1}),...]$}: Dependencies for $r_0$ stored as rejected \\
 \textit{depth = [p, $f_{depth}$]}: Minimum depth and its factor \\
 \textit{orphan = $f_{orphan}$}: Orphan factor (default value = 1) \\
 \textit{property = [[$p_0, v_0, f_0$],...]}: Properties, values and factors\\
  \textit{$D_p$= []}:  Set of proposed dependencies  for $r_0$
  
 \begin{algorithmic}[1]
  \STATE $ D_p.combine(references(r_0)+duplicates(r_0)) $
  \FORALL{$d_p$ in $D_p[]$}
     \IF {($d_p$ member\_of $D_0$) \textbf{OR} ($d_p$ member\_of  $D'_0$)}
        \STATE $D_p.delete(d_p) $
     \ELSE
     \IF {$r_0$.distance($d_p.r_p$) $> p$}
        \STATE $d_p.score.multiply(f_{depth}) $
     \ENDIF    
     \IF {$d_p.r_p.orphan()$}
        \STATE $d_p.score.multiply(f_{orphan}) $
     \ENDIF    
  \FORALL{($p_i, v_i, f_i$) in $property(p,v,f)$}
    \IF {$r_p.property(p_i)$ == $v_i$}
        \STATE $d_p.score.multiply(f_i) $
     \ENDIF
  \ENDFOR
  \ENDIF    
     
  \ENDFOR
  \RETURN sort\_by\_score\_descending($D_p$)
 \end{algorithmic}
 \label{algorithm:optimize}
\end{algorithm}
\textbf{Contextualization of dependency proposals for an issue.} 
The contextualization technique consists of finding the most likely correct and useful dependency proposals, filtering out redundant proposals, and emphasizing proposals that would be difficult to find otherwise or have features that the user wishes to find (improvement to \textit{Objective2}). Contextualization emphasizes likely results with desired features, thus improving the results of the detection algorithms by combining results and using knowledge of the domain instead of tweaking the detection algorithm itself. The proposed dependencies should likely be correct and hold additional value compared to the previous state of things. For example, a user might be more interested in similar issues reported around the same time since the issues can concern the same bug that developers have encountered in different situations. Also, an erroneous proposal is not useful either, and having too many of them erodes the users' trust in the tools; therefore, already rejected proposals can be saved and filtered out from the results.  Thus, contextualization is necessary for the solution to be more useful in practice. The technique is presented in Algorithm~\ref{algorithm:optimize} and described in more detail below.

First, our solution aggregates the scores of Algorithm~\ref{algorithm:cross-det} and Algorithm~\ref{algorithm:sim-det} for a given issue $r_0$ (Line~1 in Algorithm~\ref{algorithm:optimize}). This kind of aggregation of results of different algorithms -- or ``voting'' -- is a useful tool for improving the accuracy of results and masking errors made by one of the algorithms, especially when the components base their results on different metrics \cite{myllyaho2021misbehaviour}. For example, in our case, a direct reference detected by Algorithm 1 to $r_0$ together with a high similarity score detected by Algorithm 2 implies that someone has noticed the similarity and mentioned it in the comments. On the contrary, if $r_0$ is referenced, but the similarity score is low, the comment is more likely to be about something other than perceived similarity. The aggregation used here is simply the sum of the cosine similarity ($0..1$, see Algorithm~\ref{algorithm:sim-det}) for duplicate detection and a default value ($1$) for reference detection. An additional benefit is that no dependency is proposed twice: once by Algorithm~\ref{algorithm:cross-det} and again by Algorithm~\ref{algorithm:sim-det}.

Next, the solution examines all proposals obtained (loop comprising Lines 2--18). Lines 3--4 filter out redundant proposals. The detection techniques can result in proposals of dependencies for $r_0$ that already exist in TQC's Jira or have already been rejected by users. These proposals are considered redundant as the dependencies have been considered before and do not provide additional value for the user. For simplicity, we filter out the rejected proposals.

For the remaining proposals, our solution applies two specific contextualizations (modifications of output checker in \cite{myllyaho2021misbehaviour}) developed based on the feedback of TQC's Jira users: \textit{an issue graph-based contextualization} and \textit{a property-based contextualization}. The former emphasizes the dependencies not already in proximity in the issue graph. The latter allows the users to find the kind of dependencies they believe to be most valuable to them at a given point in time. Both rely on user-defined \emph{factors} to multiply the score of the proposals meeting the user's criteria. Thus, proposals are not filtered out, but the more relevant proposals in the user's context are emphasized.

More specifically, in issue graph-based contextualization, the scores of those dependencies from $r_0$ to issues in different issue graphs or the same issue graph with a greater distance than the given minimum depth~$p$, are increased (Lines 6--8). Also, the scores of dependencies from $r_0$ to orphans are increased as a special case (Lines 9--11). This is for two reasons. First, proximity often suggests a correct similarity with no practical relevance. For example, proposing a dependency between two children of an epic issue is likely correct but adds little value as their similarity is given because of their close relation to each other through parent-child dependencies. Second, duplicate issues within proximity are relatively easy to stumble upon when browsing through the issue graph, whereas duplicate issues further down the issue graph, issues in a completely different graph, and orphan issues are more challenging to find. Thus, proposed dependencies to issues further away from $r_0$ are considered more valuable in practice.

Meanwhile, property-based contextualization increases the score when the properties of an issue in a proposed dependency have the same values as specified by the user. This could mean, for example, environment, project, or creation time (Lines 12--16). For example, if a user wishes to find duplicates from the Qt Framework project, the scores of those proposals that have the issues in this project are increased. Applying a factor smaller than one decreases the score, resulting in a negation, such as emphasizing dependencies to other projects. 
The user is considered an expert of the product they are developing -- property-based contextualization allows them to use their experience and expertise and use the tool more flexibly.

\begin{algorithm}[t]
 \caption{CheckConsistencyAndDiagnose($r_0, G_0$)}
 \textit{$G_0 = (R_0,D_0)$ }: Issue graph for $r_0$  \\ 
 \textit{$D_i$} : Inconsistent dependencies\\
 \textit{$diag_d$} : Dependency diagnosis\\
 \textit{$diag_i$} : Issue diagnosis
 \begin{algorithmic}[1]
\STATE mergeDuplicates($G_0$)
\FORALL{$d$ in $D_0$}
    \IF {inconsistent($d$)}
        \STATE $D_i$.add($d$)
    \ENDIF
 \ENDFOR
    \IF {$D_i=\emptyset$}
    \STATE return(`Consistent') 
 \ELSE
 \STATE $diag_d$ = FastDiag($r_0, D_0$, sortByPriority($R_0-r_0$))
 \STATE $diag_i$ = FastDiag($r_0$, sortByPriority($R_0-r_0), D_0$)
 \STATE return(`Inconsistent', $D_i, diag_d, diag_i$)
 \ENDIF
 \end{algorithmic}
 \label{algorithm:cc-diag}
\end{algorithm}

\textbf{Automated consistency check and diagnosis of inconsistencies.}
Dependencies between issues need to be considered when analyzing the correctness of release assignments or priorities in issue graphs. The existing release planning models (cf. \cite{Svahnberg2010, Ameller2016}) are techniques for the task of finding an optimal release assignment from existing requirements by assigning requirements to releases. However, the other approach to release planning \add{\cite{GRuhe2005}} adopted here and at TQC is that the release assignments are manual, and the resulting release assignments are then checked for consistency  (\textit{Objective3}). That is, when an issue graph is represented in a machine-understandable manner, a consistency check is an elementary operation that can be automated. In addition, a \emph{diagnosis} can identify minimal conflict sets that lead to consistency.  The original diagnosis algorithm HSDAG (Hitting Set Directed Acyclic Graph)~\cite{Reiter1987} uses breadth-first search to find all minimal sets of constraints that could be deleted to restore consistency. Several improved diagnosis algorithms have been developed~\cite{felfernig2014conflict}. However, clearly defined dependency types (e.g.,~\cite{Carlshamre2001,Dahlstedt2005,Felfernig2018}) form the basis for any automation, and semantics can have a few alternative interpretations based on context. For example, the `requires' dependency can be interpreted so that the required issue must be in an earlier release or can be in the same release.
\begin{figure*}[t]
	\centering
		\includegraphics[width=0.95\textwidth]{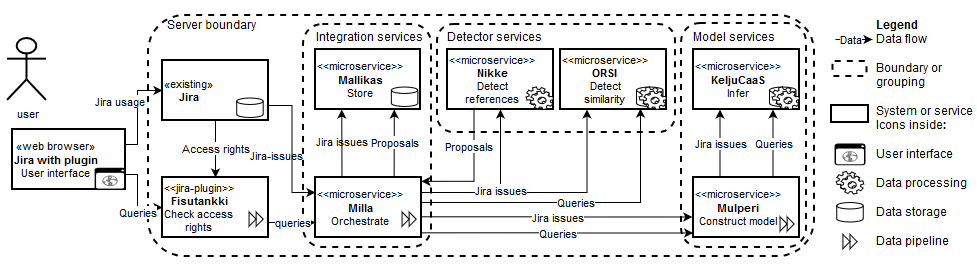}
	\caption{The software architecture of the artifact. }
	\label{fig:architecture}
\end{figure*}

In our solution for consistency check and diagnosis, we utilize `requires' and `parent-child' dependencies, which have well-defined semantics  that take priorities and release assignments into account; the details in the TQC context are described in Drawback~4 in Section~\ref{section:currentDrawbacks}. In addition, our solution merges issues with the `duplicate' dependency between them, and the resulting merged issue inherits all dependencies from the merged issues; this is the first step (Line 1 of Algorithm~\ref{algorithm:cc-diag}).
The consistency check is a procedural method that evaluates, for each dependency, whether the conditions of the dependency are satisfied and reports the violated dependencies (Lines 2--6).

If the dependency contains inconsistent dependencies, diagnosis can be invoked. 
We adopted FastDiag (see details in ~\cite{Felfernig2012A}), an efficient divide-and-conquer algorithm used to determine preferred diagnoses of constraint sets. Diagnosis applies a CSP representation of an issue graph where dependencies, priorities, and releases  become the constraints of CSP. Constraints are assumed to be in a lexical order according to their priorities: a higher priority constraint is retained if possible, even if all lower priority constraints would have to be removed. 
\emph{The issue diagnosis} (Line 10) identifies a set of issues that need to be assigned to a different release or re-prioritized or removed to restore the consistency of the network. For this diagnosis, each issue is considered a constraint that can be relaxed or 'diagnosed away'.
\emph{The dependency diagnosis} (Line 11) determines a set of dependencies whose removal from the issue graph restores consistency. The idea of diagnosis is that a user is then presented with two alternative solutions that lead to a consistent release assignment. Rather than automated change, the user must decide whether to act upon either diagnosis or find alternative actions to resolve the inconsistency.

\section{RQ3: Artifact implementation}
\label{section:implementation}

\begin{figure*}
    \centering
    \begin{minipage}{0.755\textwidth}
        \centering
			\fbox{\includegraphics[width=\columnwidth]{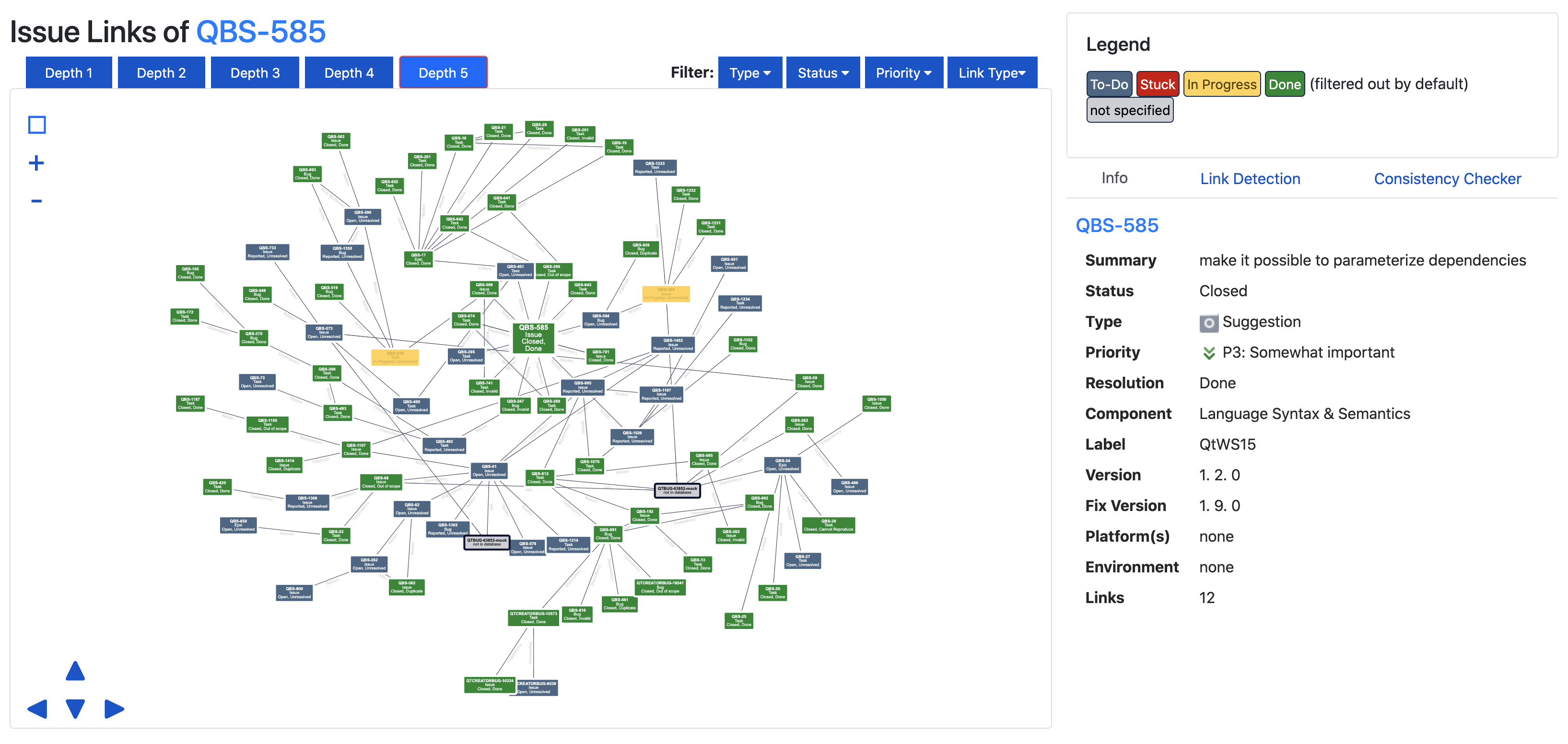}}
    \end{minipage}\hfill
    \begin{minipage}{0.245\textwidth}
        \centering
			\fbox{\includegraphics[width=\columnwidth]{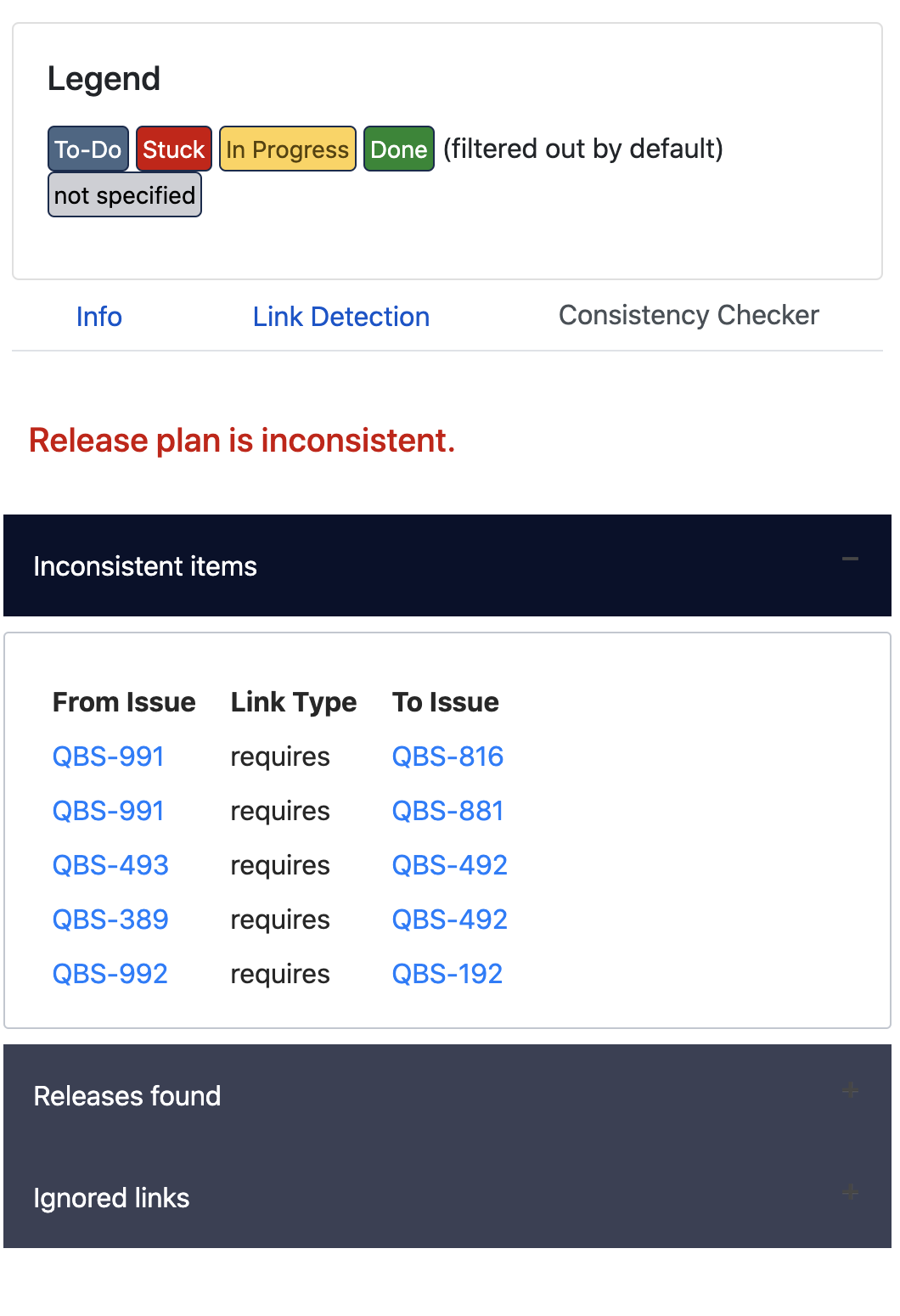}}
    \end{minipage}
		    \caption{Two screen captures of Issue Link Map for the issue QBS-585 of TQC Jira. The p-depth issue graph to depth five ($G^5_0$) and the properties of the issue ($r_0$) from Jira are shown on the left. The consistency check tab is shown on the right.}
    \label{fig:screenshot}
\end{figure*}   

This section describes the developed artifact (cf. Section~\ref{section:researchMethod}). We first elaborate on the objectives of the artifacts  and then describe the implementation.

\subsection{Artifact design objectives}
\label{sec:artifactobjectives}
We articulate the design objectives using the eight ISO25010 quality model characteristics\cite{ISO25010}. 
\begin{itemize} 
    \item \emph{Functional suitability}. The artifact needs to implement the techniques described in the previous section.
    \item \emph{Performance efficiency}. The artifact needs to handle a large number of issues efficiently, and the efficiency of Jira may not be unacceptably damaged. As an illustration, TQC's Jira users estimated this goal as ``responses, even to the largest requests, within a few seconds''.
    \item \emph{Compatibility}. The artifact itself needs to be compatible (co-exist and interoperate) with Jira's functionality and data without the need to develop additional software for interchanging data or accessing functions.
    \item \emph{Usability}. The usage of the artifact needs to  integrate smoothly with Jira and the present way of working. 
    \item \emph{Reliability}. Integrating the artifact and its data should not interfere with Jira's current issue management. This was clearly stated by TQC's Jira users, who had avoiding any risk concerning their current Jira management as their top priority.
    \item \emph{Security}. The solution must not compromise private data and must adhere to the company access policies.
    \item \emph{Maintainability}. The architecture needs to support easy evolution and extension as Jira evolves, and allow for easy integration of new techniques.
    \item \emph{Portability}. The solution should not be strongly tied to any particular technology other than Jira or impose unnecessary additional installation decisions.
\end{itemize}

\subsection{Artifact design}
\label{sec:artifact-design}

We implemented the artifact as a Jira plugin and service-based system consisting of independent microservices (${\rightarrow}$maintainability, compatibility), which in practice operate in a choreographic manner following a layered architectural style. The services collaborate through JSON-based messages  following an earlier defined generic ontology~\cite{Quer2018} that adheres to REST principles (${\rightarrow}$portability). There are three classes of microservices and the plugin as summarized below and in the architecture diagram in Figure~\ref{fig:architecture}.

{\bf 1. Integration microservices.} First, one microservice (\emph{Milla}) integrates with Jira, fetches issue data, and constructs dependencies as separate, first-class entities. We realized the integration using Jira's existing OAuth-based REST API (${\rightarrow}$portability, security). A full projection of TQC's Jira issues is made, and relevant information is cached to provide more efficient access to issue data (${\rightarrow}$efficiency). The resulting issue and dependency data from Jira are cached in a local database embedded into an auxiliary integration microservice (\emph{Mallikas}). Frequent updates fetch new and changed issues from TQC's Jira (${\rightarrow}$compatibility). 

{\bf 2. Detector microservices.} After completing the data projection, the integration service (\emph{Milla}) sends the resulting issues and dependencies -- or their changes when updating -- to the detector microservices for processing  (${\rightarrow}$efficiency). 
The reference detector (\emph{Nikke}) searches for missing dependencies (i.e., implementing Algorithm~\ref{algorithm:cross-det} presented in Section~\ref{section:drawbacksManagement}) and the similarity detector (\emph{ORSI}) searches for duplicated issues (implementing Algorithm~\ref{algorithm:sim-det}). 
The reference detector \emph{(Nikke)} returns proposed dependencies (`proposals' in Figure~\ref{fig:architecture}), which are then stored in the same local database (\textit{Mallikas}) as the existing dependencies, applying the `proposed' value for the status-property. 
However, the similarity detector (\emph{ORSI}) requires persistence on the service side to optimize the similarity due to clustering and vector-based algorithms. Therefore, the proposals are stored internally in the cluster (${\rightarrow}$efficiency).

{\bf 3. Model microservices.} The integration service (\emph{Milla}) also sends the issue graph~($G$) to the model microservices (\emph{Mulperi} and \emph{KeljuCaas}). These microservices translate the issue graph into a more general knowledge representation and store the data as a map datatype with issues as keys and a list of said issues' neighbors along with the corresponding dependency types. The way in which the issue graphs are stored allows easy extraction of various $p$-depth issue graphs ($G^p_0$) by following the dependencies recursively to the required depth (${\rightarrow}$efficiency).

\begin{table*}[t]
\centering
\caption{A summary of evaluation, metrics, and used data-sets.}
\label{tab:func-eval-metrics}
\begin{tabularx}{\textwidth}{p{2.3cm} p{4cm} X p{2.6cm}} % NOTE: Change multirows if you change p-values
\toprule
\textbf{Technique}     & \textbf{Metric Id}  & \textbf{Description}  & \textbf{Data-sets} \\ \midrule
\multirow{2}{2.3cm}{Issue graph handling} & \#dependencies & Number of dependencies in each issue & Qt Repository      \\
& \#p-depth-graphs & Number of $p$-depth issue graphs\\
& \#issues-in-p-graphs & Number of issues in  $p$-depth issue graphs\\

\midrule

\multirow{2}{2.3cm}{Dependency \newline detection}  & \#issues           & Number of issues for which a dependency is proposed & \multirow{2}{2.6cm}{Qt Repository and \newline Duplicate set \#1  and \newline Duplicate set \#2} \\
& \#proposals & Number of dependency proposals &         \\
& \#depth-3-distance & Number of issues with more than 3 edges apart& .        \\ %\rule{0pt}{3ex} 
\cline{2-4}
&  accuracy, precision, recall, \newline F-measure & Quality classification metrics based on a cross-validation analysis of detectors dependency predictions & Cross-validation set  \\
\midrule

\multirow{2}{2.3cm}{Consistency check  and diagnosis} & \#requires-inconsistent     &  Number of inconsistent \emph{requires} dependencies  & Qt Repository      \\
& \#parent-child-inconsistent     & Number of inconsistent \emph{parent-child} dependencies  & \\                    
& \#p-depth-consistency &  Number of consistent $p$-depth issue graphs  &        \\ 
& \#issue-diagnosis-count & Number of issues  diagnosed to be removed \\
& \#dependency-diagnosis-count &  Number of dependencies  diagnosed to be removed\\
& issue-diagnosis-success & Success of issue diagnosis \\
& dependency-diagnosis-success & Success of dependency diagnosis \\
                               
\bottomrule
\end{tabularx}
\end{table*}

{\bf 4. User interface plugin.} Users interact through a dedicated Jira plugin (\textit{Fisutankki}) installed in TQC's Jira. The plugin technology integrates the user interface into Jira and Jira's security mechanisms (${\rightarrow}$usability, compatibility). This allows public access, where authenticated users adhere to Jira's security schema (${\rightarrow}$security).

On the users' side, \emph{Issue Link Map}~\cite{Lueders2019} (Figure~\ref{fig:screenshot}) is embedded in the Jira plugin (\textit{Fisutankki}), which creates a browser-based user interface (${\rightarrow}$usability, compatibility). 
A central part of the user interface is a 2D representation of a $p$-depth issue graph ($G^p_0$). The issue ($r_0$) in focus is in the center, and the other issues are automatically positioned around it circularly, depending on their depth. The user can select the desired depth, up to depth five, from the top-left, rearrange the issues, zoom in and out, etc. The colors indicate the status of the issues. A set of filters, such as type or status, can be applied to the visualization.
On the right, various tabs represent the other techniques. The first tab shows the basic information for the selected issue as in Jira because the 2D diagram cannot convey all the details of an issue. The second tab shows dependency proposals, which are then also shown in the 2D diagram as dashed lines. The third tab shows the results of the consistency check. 

The user interface design was periodically shown and discussed with two TQC product managers and optimized according to their feedback. The user interface primarily tackles \textit{Drawback1}, as users can view connected issues in total, and offers an easy way to see and accept dependencies. 

The user interface accesses the functionality provided by other services through REST calls, which we refer to as \emph{queries} in Figure~\ref{fig:architecture}. Each query goes through the plugin (\textit{Fisutankki}) that applies Jira's security policies. Then the integration microservice (\emph{Milla}) orchestrates all queries to other microservices (${\rightarrow}$maintainability). The elementary functionality to initiate the user interface is to query an issue graph to depth five ($G^5_0$) from the model microservices, which the user interface visualizes to the desired depth. 

The integration microservice (\textit{Milla}) processes a user's query for a dependency proposal implementing Algorithm \ref{algorithm:optimize}. First, it combines reference proposals (in \textit{Mallikas}), similarity proposals (in \textit{ORSI}) and removes rejected proposals (stored in \textit{Mallikas}). Second, it calls the model services for the desired p-depth issue graph ($G^p_0$) to apply the issue graph-based contextualization. Third, it queries the cached data (in \textit{Mallikas}) for the property-based contextualization. A user can accept a proposed dependency that requires them to specify its type, or reject or disregard the proposal. Provided that the authorized user has sufficient privileges, the plugin (\textit{Fisutankki}) writes accepted decisions to Jira as new dependencies, while the local database (\emph{Mallikas}) stores rejection decisions.

The integration service (\textit{Milla}) forwards a user's query for consistency check and diagnosis to the model services that first construct an issue graph ($G^p_0$) internally and prepare data for inference, such as translating the version numbers to integers (in \textit{Mulperi}). 
Then the consistency check is carried out, and, in case of inconsistency, the issue graph is read to constraint programming objects, and the Choco solver~\cite{choco} (in \textit{KeljuCaaS}) is used to infer diagnosis (Algorithm~\ref{algorithm:cross-det}). 

The microservices are deployed to the same server as TQC's Jira, which relies on the server's security mechanisms (`server boundary' in  Figure~\ref{fig:architecture}). Although the microservices use secure communication, the data is not transferred to other servers remaining behind the server's firewall -- only the plugin's (\textit{Fisutankki}) REST endpoint is publicly accessible (${\rightarrow}$security).

\section{Evaluation}
\label{section:quality}

We evaluate the eight ISO25010 quality model characteristics in three different manners. For five of them, we argue their realization by design as described in the
previous section: compatibility, usability, security, maintainability, and portability.  For a sixth one, reliability, we did not encounter problems during the test period. We experimented with various test setups, and the final tests took over a week without discontinuity of service. A small number of performance tests behaved abnormally, such as 9 out of 119,920 (0.0075\%) dependency queries, which should take roughly the same time but in practice took more than twice the average time. Since we could not reproduce the behavior, we assume that they were caused by the infrastructure, such as Java's garbage collection. Although it was only experimentally used for several months, the system was  operational in TQC’s test Jira without discontinuity in the service.

The two remaining characteristics, functional suitability and performance efficiency, were subject to dedicated evaluation. 
More precisely, we carried out verification of microservices by system tests for functionality (Sections \ref{sec:analysisOfBackgroudFeratures}-\ref{sec:ConsistencyManagementVerification}) and performance (Section~\ref{sec:performanceEvaluation}), and validation of the solution by user interviews (Section~\ref{sec:validation}).
Table~\ref{tab:func-eval-metrics} summarizes the metrics for functionality, and we measured performance by the execution times. 

Throughout the evaluation, we had our artifact deployed to the TQC's test environment (cf. Section~\ref{sec:ReseachContextTQC}), and we used its full public data -- referred to as \emph{Qt Repository} -- consisting of 119,920 issues in 20 different projects and their 29,582 dependencies. % from TQC's Jira. 
Additionally, we executed all microservice verification tests for comparability using the same Linux virtual computer with a single Intel Xeon CPU E7-8890 v4 2.20GHz processor and 50GB memory at University of Helsinki, Finland.

\subsection{Evaluation of baseline techniques} 
\label{sec:analysisOfBackgroudFeratures}

\textbf{Evaluation design} The  evaluation goal of the baseline techniques was to verify the functionality  (Section~\ref{sec:drawbacksManagement-background}) and  better understand the characteristics of TQC's Jira data with respect to existing dependencies. We first carried out the evaluation as an  exploratory analysis as a set of batch process measures. We enumerated different projects, and then we calculated in total and in different projects the number of different types of issues, the number of different types of dependencies and whether the dependencies were inter-project and cross-project dependencies, and the number and types of dependencies from each issue. Finally, we enumerated all possible $p$-depth issue graphs, for which we calculated the total number of issues and the number of issues at each level. In addition to these batch process measures, we manually inspected selected issues, such as the ones with the highest number of dependencies. Below, we report the evaluation results of metrics related to the topology and size of the generated $p$-depth issue graphs, as shown by the first block in Table~\ref{tab:func-eval-metrics}. 

\textbf{Data-sets.} \textit{QT}: Qt Repository, all issues, and their dependencies.

\textbf{Evaluation results.}
In total, 31,182 issues~(26\%) have at least one dependency declared by TQC's Jira users by Issue Links in Jira (\textit{\#dependencies} in Table~\ref{tab:func-eval-metrics}), meaning that~88,738 issues~(74\%) are orphans before any automated dependency detection. Out of the issues that have dependencies,~75\%  have only one dependency. The average is 1.7, and the median is 1. As noted in Section~\ref{sec:ReseachContextTQC}, issues are sometimes used for grouping, resulting in and explaining that the maximum number of dependencies is~139, and~24 issues have at least~50 dependencies.
Generating all different p-depth issue graphs for all issues (i.e. $\forall r_i \in R$ we generated a $G^p_i$ $\forall p \in [1,n]$ so that $G^n_i$ = $G_i$) resulted in~320,159 issue graphs (\textit{\#$p$-depth-graphs}). 
By analyzing the number of issues in various $p$-depth issue graphs (\textit{\#issues-in-$p$-graphs}), we observed that the largest issue graph consists of~8,952 issues, and the maximum depth in its topology is~42. This issue graph is exceptionally large, with many subgraphs, as the next largest maximal issue graph consists of~162 issues with a maximum depth of~16. Finally, we inspected the number of issues in all different $p$-depth issue graphs (\#issues-in-p-graphs) and observed high variance and exponential growth in the number of issues at low depths. For instance, 5-depth graphs have a minimum of 5, an average of 210.5, and a maximum of 1778 issues.

\textbf{Summary.}
This exploratory analysis of the issue dependency network ($G$) reveals that there are many dependencies but also many disjoint issue graphs ($G_0$), including orphans. The number of issues in $p$-depth issue graphs can often be quite large and grow rapidly and exponentially as a consequence of average dependency count but also the grouping issues in the topology.  
In practice, issue graphs up to depth five are still meaningful for a user, but typically, issue graphs at greater depths contain too many issues and dependencies.

\subsection{Evaluation of dependency management techniques}
\label{sec:DependencyManagementVerification}

\textbf{Evaluation design.} The goal of dependency management evaluation was to assess the validity and coverage of the detectors. We applied reference detection (Nikke) and duplicate detection (ORSI) to each issue of data-sets \textit{QT}, \textit{D1}, and \textit{D2} introduced below. We also differentiated the union and  intersection of the results to analyze dependencies that both detectors or only one detector proposed, respectively.
Statistical quality analysis with data-set \textit{CV} provides cross-validation with \textit{k=10}. The metrics are in the second block in Table~\ref{tab:func-eval-metrics}. 

\textbf{Data-sets.} The analysis was carried out for each issue in the following data-sets.

\begin{itemize}
    \item \textit{QT}: Qt Repository, all issues, and their dependencies. 
    \item \textit{DS1}: Duplicate set \#1, a sub-set of \textit{QT} consisting of all 5,839 issues marked as duplicates without `duplicate' dependency (See Drawback 3 in Section~\ref{section:currentDrawbacks}). As these issues were duplicates, we assumed a duplicating issue in \textit{QT}.
    \item \textit{DS2}: Duplicate set \#2, a sub-set of \textit{DS1} consisting of all 914 issues resolved as duplicates but without any dependencies. 
    \item \textit{CV}: Cross-validation set, a sub-set of 2,936 pairs of issues without existing dependencies in \textit{QT} structured as follows. One group consisted of 1,437 pairs of issues reported by TQC domain experts as duplicates in TQC's Jira, meaning that one issue of each pair was marked as a duplicate, and each pair had a `duplicate' dependency between them. We labeled this first subset of \textit{CV} as \textit{duplicates}. To generate a balanced data-set, we used another group of 1,499 pairs of randomly selected closed issues with no duplicate resolution reported in TQC's Jira that we labeled as the \textit{not-duplicates} sub-set of \textit{CV}. 
\end{itemize}

\textbf{Evaluation results}
The results of the quantitative analysis for the three first data-sets are shown in Table~\ref{tab:dep-det-results}.
An analysis of distribution reveals that the duplicate detection typically proposes several dependencies to all issues. In contrast, in the reference detection,  most issues have only one proposal, and a few issues have several proposals as a list or table  (cf. Section~\ref{sec:ReseachContextTQC}).
In the case of issue graph-based contextualization, only 2\% of the proposals were three edges apart or closer (\textit{\#depth-3-distance}) in Qt Repository, and all resulted from duplicate detection (\emph{ORSI}).

Table~\ref{tab:dep-det-results-crossvalidation} shows the results of the cross-validation analysis for detector services using the \textit{CV} data-set. We compare both detectors, although reference detection is not designed only for duplicate detection, and therefore the results must be interpreted with this in mind. The low recall is expected for reference detection \textit{(Nikke)}, but high precision is not expected. In order to verify the results, we decided to analyze handpicked sample issues in which reference detection found a dependency by reading through the text and comments of the issues. The analysis verified the results, and we discontinued verification after about 30 checks, which were all correct. The verification also revealed that it is customary to add a comment to an issue about duplication, as most had a comment about duplication, and the rest noted duplication in the description field, which explains the high precision. 
For duplicate detection (\textit{ORSI}), the optimal similarity threshold value (see Algorithm~\ref{algorithm:cross-det}) was reached through a set of experiments by fine-tuning the similarity threshold with $\pm$0.1 deviations until it reached the global maximum for F-measure.
% Duplicate detection reports balanced quality metrics, with special emphasis on high precision. 
Compared with the data in Table \ref{tab:dep-det-results}, our solution tries to reduce false positive instances as much as possible, given the large number of issues and, as a consequence, the large number of dependency proposals. This idea is reinforced if compared with reference detection results, where perfect precision is achieved.

\begin{table}[t]
\caption{The results of dependency detection in terms of \textit{\#issues}  and \textit{\#proposals} as defined in Table~\ref{tab:func-eval-metrics}}
\centering
\label{tab:dep-det-results}
\begin{tabularx}{\columnwidth}
{p{2cm} 
p{2.45cm} 
>{\raggedleft\arraybackslash}p{0.7cm}  >{\raggedleft\arraybackslash}p{0.5cm} 
>{\raggedleft\arraybackslash}X}

\toprule
\textbf{Data-set} & \textbf{Detector} & \textbf{\textit{\#issues}} &\textbf{(\%)}& \textbf{\textit{\#proposals}} \\ \midrule
Qt Repository     & Reference detection              & 24,097 &(20\%)    & 31,646     \\
(\textit{Qt} data-set)                & Duplicate detection               & 45,570 &(38\%)    & 578,739     \\
                  & Union  & 60,250 &(50\%)    & 610,348     \\
                  & Intersection & 1,727 & (1\%)& 1,801  \\
                  \midrule
Duplicate set \#1       & Reference detection              & 3,275  &(56\%)     &3,935
      \\
(\textit{DS1} data-set)                  & Duplicate detection               & 2,479  &(45\%)     & 33,153     \\
                  & Union  & 4,457  &(76\%)     & 37,208
     \\ 
                     & Intersection & 377
 &(6\%) & 388
  \\
                     
                     \midrule
Duplicate set \#2       & Reference detection              & 182    &(20\%)    & 208     \\
(\textit{DS2}  data-set)                  & Duplicate detection               & 423    &(46\%)    & 5,526     \\
                  & Union  & 526    &(58\%)     & 5,742     \\ 
              & Intersection & 15 & (2\%) & 16  \\\bottomrule
\end{tabularx}
\end{table}

\begin{table}[t]
\caption{Cross validation results of detectors for the \textit{CV} data-set.}
\centering
\label{tab:dep-det-results-crossvalidation}
\begin{tabularx}{\columnwidth}{p{1.5cm} L L }									\toprule
\textbf{Measure} & \textbf{Reference detection} & \textbf{Duplicate detection\ } \\ 
\hline

Accuracy	&77.15\% & 91.66\%\\
Recall	&53.31\% & 86.15\%\\
Precision	&100.00\% & 96.42\%\\
F-measure	&69.54\% & 91.00\%\\

 \bottomrule
\end{tabularx}

\end{table}

\textbf{Summary.}
The quantitative analysis shows that the detectors have the potential to expand the issue dependency network  by proposing a significant number of  dependencies.
The number of issues for which reference detection makes proposals is relatively large, but the number of dependencies for one issue is small --  typically one and even on average 1.4 proposals for issues for which a proposal is made.  
In contrast, duplicate detection finds proposals for many issues and results in many proposals per issue, especially considering that the proposals are about duplicates: 38\% of issues cannot be duplicates but the results include false positives. Likewise, the number of issues in Qt Repository (119,920),  compared to the number of proposed dependencies (578,739), indicates false positives. Only a small number of false positives can be explained by closely connected issues, such as between the children of an epic based on issue graph-based contextualization. However, as the underlying principles of detectors are different, the number of proposals is not surprising.  The small intersection of proposals for Qt repository shows that the detectors complement each other, while the larger intersection for duplicate sets indicates that detectors can also support each other. The precision and the small number of proposals of reference detection justify its default score of 1.0, while duplicate detection itself provides a score. Contextualization relying on the score-based approach seems appropriate to combine, prioritize, and filter relevant proposals for users.

\subsection{Evaluation  of consistency check and diagnosis}
\label{sec:ConsistencyManagementVerification}

	\begin{table*}[t]																					
	\caption{A summary of consistency check and diagnosis results until depth of 10 ($G^1_i$...$G^{10}_i$).} %\mikko{Generated from excel, do not change here}}																					
	\label{tab:cc_and_diag_results}		\centering																			
	\def\arraystretch{1.2}																					
	\begin{tabularx}{\textwidth}{p{5cm} L L L L L L L L L L}																					
	\toprule																					
	&\multicolumn{10}{c}{\textbf{Depth}}\\																					
	\textbf{Measure}	&	\textbf{1}	&	\textbf{2}	&	\textbf{3}	&	\textbf{4}	&	\textbf{5}	&	\textbf{6}	&	\textbf{7}	&	\textbf{8}	&	\textbf{9}	&	\textbf{10}	\\
	\hline																					
	\#requires-inconsistent average	&	0.7	&	3.7	&	5.1	&	8.7	&	14.9	&	24.4	&	36.1	&	45.6	&	55.2	&	67.9	\\
	\#parent-child-inconsistent average	&	0.8	&	4.1	&	4.4	&	6.4	&	12.1	&	20.7	&	32.4	&	36.8	&	34.6	&	36.9	\\
	\#p-depth-consistency (\%)	&	93\%	&	72\%	&	49\%	&	30\%	&	20\%	&	13\%	&	7\%	&	4\%	&	3\%	&	2\%	\\
%	\% Inconsistent	&	6.9	&	27.6	&	50.7	&	69.8	&	80.4	&	87.1	&	92.7	&	95.6	&	97.2	&	97.9	\\
	\#issue-diagnosis-count average	&	1.1	&	1.7	&	3.0	&	4.6	&	7.2	&	7.3	&	8.6	&	9.2	&	9.5	&	10.2	\\
	issue-diagnosis-success\textsuperscript{1} (\%)	&	100\%	&	100\%	&	100\%	&	99\%	&	91\%	&	69\%	&	54\%	&	39\%	&	28\%	&	21\%	\\
	\#dependency-diagnosis-count average	&	1.5	&	7.8	&	9.4	&	14.9	&	25.6	&	33.7	&	38.8	&	48.0	&	51.2	&	57.5	\\
	dependency-diagnosis-success\textsuperscript{1} (\%)	&	100\%	&	100\%	&	100\%	&	100\%	&	98\%	&	80\%	&	67\%	&	55\%	&	41\%	&	32\%	\\
	\bottomrule																					
	\hline																					
	\end{tabularx}																					\textsuperscript{1} Success is measured by not exceeding the time limit (5 seconds) since all other diagnoses found a solution. 
	\end{table*}											

\textbf{Evaluation design.}
The goal of the evaluation was to analyze the consistency in TQC's Jira as well as to verify the technical feasibility of consistency check and diagnoses.
We analyzed the consistency of all `requires' and `parent-child' dependencies individually, i.e., taking into account only the dependency and the issues on both ends without any other dependencies of the issues and the consistency and diagnosis of all $p$-depth issue graphs ($G^p_0$). The metrics are outlined in the third block of Table~\ref{tab:func-eval-metrics}. Since we noticed that different Jira projects do not have comparable and machine-understandable version numbering, we disregarded all cross-project dependencies from the analysis. As diagnoses are computationally heavy operations, we set the time limit to five seconds for each $p$-depth issue graph and did not carry out the diagnoses to any greater depth. A five-second limit was considered reasonable from the user's perspective. This limitation was also necessary as the tests already took over a week, and a larger limit or removing a limit would have required a significantly longer time or design change with little practical value. 

\textbf{Data-sets.} The analysis was carried out for the following data-sets.

\begin{itemize}
    \item \textit{QT}: Qt Repository, all issues, and their dependencies
    \item \textit{Deps}: Dependency set, a sub-set of \textit{Qt} consisting of 3,989 `requires' and 8,222 `parent-child' inter-project dependencies and the issues in both ends of each dependency.
\end{itemize} 

\textbf{Evaluation  results}
The consistency check for each dependency individually for the \textit{Deps} data-set found inconsistency in 780/3,989 (20\%) of `requires' dependencies (\textit{\#requires-inconsistent}) and 884/8,222 (11\%) of `parent-child' dependencies (\textit{\#parent-child-inconsistent}).
The results of consistency check and diagnoses for all 320,159 $p$-depth issue graphs in the \textit{QT} data-set are summarized in Table~\ref{tab:cc_and_diag_results} by depth to a depth of 10 ($G^1_i$...$G^{10}_i$), to draw an overview on the evolution of inconsistencies with issue graph depth. 
For issue graph sizes, the first unsuccessful and the last successful execution of issue diagnosis were carried out for the issue graphs of sizes 371 and 701 issues, respectively. The respective numbers for the dependency diagnosis were 580 and~1362. 

\textbf{Summary.}
In the case of consistency check, we observe that a significant amount (11-20\%) of all dependencies are inconsistent. However, some  inconsistencies result from new issues that have not yet been assigned to a release. Inconsistency  becomes prevalent for issue graphs at any greater depth, as shown by the decreasing \emph{\#$p$-depth-consistency}, presented as a percentage in Table~\ref{tab:cc_and_diag_results} (the 3rd row). 
Moreover, the number of detected inconsistencies increases significantly with greater depths of issue graphs.
There are already dozens of inconsistencies at relatively small depths, as shown by the two first rows of Table~\ref{tab:cc_and_diag_results}. Consequently, from the practical perspective, complete consistency remains an elusive target, and the analysis context of the consistency check is practically relevant for a user until depth five.

Regarding the diagnosis, the diagnoses start to fail from depth 4, i.e., take more than five seconds, and the success rate falls quite rapidly at any greater depth (\textit{issue-diagnosis-success (\%)} and \textit{dependency-diagnosis-success (\%)} in Table~\ref{tab:cc_and_diag_results}).
At small depths, when all diagnoses are successful, we see that the diagnosis of dependencies essentially proposes to remove all inconsistent dependencies (\textit{\#dependency-diagnosis-count
= \#requires-inconsistent + \#parent-child-inconsistent}) while the diagnosis of issues requires changes to the priority or release of a significantly smaller number of issues (\textit{\#issue-diagnosis-count}).  
The relatively small increase in these numbers as depth increases means that only the smallest issue graphs are diagnosed successfully -- there is a significant variance in the issue graph sizes at greater depths, as covered above. A qualitative analysis of diagnosis results revealed that lexical order does not always work properly when dependencies are not clearly prioritized, and issues appear in a few priority classes.
Consequently, the evaluation shows that the implemented diagnoses are functionally feasible but, for a user, computationally meaningful until issue graphs containing less than 1000 issues and algorithms should provide alternative diagnoses.

\subsection{Performance evaluation}
\label{sec:performanceEvaluation}

\textbf{Evaluation design.}
The goal of performance evaluation was  to assess the efficiency of all functionality with respect to required computing time and, in particular, give fast enough responses to users (cf. Section~\ref{sec:artifactobjectives}). We divided the performance evaluation into (\emph{i}) batch tasks covering initial processing and updates, and (\emph{ii}) queries, which are scenarios for a user.  
In order to individually evaluate batch tasks, we divided the performance evaluation into a data projection from Jira, which also covers processing dependencies, and processing in both detectors.  
We report the average times of five tests to eliminate random errors. 
For the evaluation of the queries, we applied the various usage scenarios to microservices as orchestrated end-to-end systems, measuring the time from sending a user's query request to a response. This corresponds with the time for submitting a query to and getting a response from the integration service (\emph{Milla} in Figure~\ref{fig:architecture}). Since we focus on the microservices, we omitted user interface rendering and Jira plugin functionality. 
We analyzed execution times in the data-sets for dependency query for all issues, and issue graph initialization, consistency check, and diagnosis for all $p$-depth issue graphs.

\textbf{Evaluation data-sets.}
We applied various data-sets for evaluation, as detailed below.

\begin{itemize}
    \item \textit{Qt Repository}. All issues and their dependencies.
    \item \textit{Large issue graphs}. A sub-set of \textit{Qt Repository} containing all $p$-depth issue graphs for any $p$ with at least 8,000 issues, which integrate 82,640 different issue graphs. We use this data-set for the worst-case scenario. 
    \item \textit{Sizeable issue graphs}. A sub-set of \textit{Qt Repository} containing all $p$-depth issue graphs for any $p$ with 500-1,000 issues, which integrates 14,783 different issue graphs. We use this data-set to represent a possible large case scenario that a user might be interested in, being similar with the largest 5-depth issue graphs.
    \item \textit{Update data-set}. The small project (QTWB) as a sub-set of \textit{Qt Repository} consisting of 27 issues and 9 dependencies to simulate an update. This data was first manually removed from \textit{Qt Repository} and our system.
\end{itemize}

\textbf{Evaluation results.}
The results of the performance evaluation are summarized in Table~\ref{tab:PerformanceResults} as average execution times. 
Data transfer between servers took the majority of the time in the data projection, but even when all software is deployed to the same server, we found that data projection takes several minutes because of the large amount of data and Jira's inefficient REST interface, which requires fetching issues as sets of individual issues.
The $p$-depth issue graph queries are fast, and depend on the size of the issue graph because many issue properties are returned, making the return data large.
The execution times of dependency queries have a small variance and do not depend on data size: the minimum time was 1.3 seconds and 62 queries took over 2.5 seconds, out of which 25 queries returned fewer than 10 proposals.
The time required for the consistency check appears to increase practically linearly with respect to the number of issues. The data has minor variation as 0.15\% of queries take 10-17 seconds.
We do not present average times for diagnoses because diagnoses for large graphs were not calculated; diagnosis under a five-second limit has been discussed in the previous subsection.

\textbf{Summary.} The evaluation results show that the initial operations take hours, but they are performed as a batch process upon system initialization. Updates are then relatively fast, up to tens of seconds. 
Queries other than diagnosis are within reasonable limits for a user as they take less than five seconds on average, even for the largest issue graphs. However, the tests with \emph{Sizeable issue graphs} show that operations are fast and even diagnoses are then feasible as discussed above.
Although we did not measure the time required for authorization and visualization in the Jira plugin, we have not experienced any significant delays.

\begin{table}[t]
	\caption{Performance analysis results.}
	\label{tab:PerformanceResults}	\centering
\begin{tabularx}{\columnwidth}{p{2.3cm} p{4.2cm} L }
\textbf{Task \textit{(Data-set)}}& \textbf{Technique} &  \textbf{Time} \\
\toprule
\multirow{2}{3cm}{Data processing \newline \textit{(Qt Repository)}}
&Data projection (\emph{Milla})& 40 m\\
&Reference processing (\emph{Nikke})& 31 m \\ %30 m 32 s\\
&Similarity processing (\emph{ORSI})&4 h 34 m\\
\midrule
\multirow{2}{3cm}{Update processing \textit{(Update data-set)}}
&Data update projection (\emph{Milla})& 4.4 s\\
&Reference processing (\emph{Nikke})& 1.4 s\\
&Similarity processing (\emph{ORSI})& 28.6 s \\
\midrule
\multirow{2}{3cm}{Queries \newline \textit{(Qt Repository})}
&$p$-depth issue graph query  &  0.3 s\\
&Dependency query & 1.7 s\\
&Consistency check query & 1.9 s \\
&Diagnosis & --- \\
\hline 
\multirow{2}{3cm}{Queries \newline \textit{(Large issue graphs)}}
&$p$-depth issue graph query &0.7 s\\
&Consistency check query   & 4.7 s \\
\hline
\multirow{2}{3cm}{Queries \newline \textit{(Sizeable issue graphs)}} % 500-1000
&$p$-depth issue graph query & 0.01 s\\
&Consistency check query   & 0.2 s \\

\bottomrule
\end{tabularx}
\end{table}

\subsection{Validation interview study}
\label{sec:validation}

\textbf{Validation study design}
The goal of validation was to assess whether the user considered the techniques valuable. That is, in addition to the iterative approach of constant feedback from TQC, we carried out the final validation of the artifact by interviewing five of TQC's Jira users, who were all active Jira users who tested and used our solution but had not yet been involved in the design and implementation process:
two release managers, one software architect, one product manager, and one developer. We interviewed each respondent individually, following a semi-structured approach.  Three researchers carried out the interviews in Finnish, following  predefined roles: one researcher acted as the leading interviewer, and others took notes and asked clarification questions. We carried out one interview first and four interviews on another day. The main interviewer was the same in all interviews, while the other two interviewers changed. Each interview took about 1.5 hours.
We  instructed each respondent  to use the system beforehand, allowing one week for this. During the interviews, the respondents were asked to use a shared meeting room monitor to demonstrate and explain the tasks while interviewers voice-recorded and took notes of the process.  
The structure of the interviews consisted of an introduction, the concept and visualization of issue dependencies, consistency check and diagnosis, dependency proposals, and data updates.
We had prepared and printed a set of slides (available in GitHub\footnote{The interview questions: \href{https://github.com/ESE-UH}{https://github.com/ESE-UH})} explicating the themes and the questions in the themes, but only some example screenshots and diagrams were shown to the respondents on paper when needed to explain something.  The  slides contained the questions to which the interviewers sought answers rather than questions for respondents, adhering to the principles of a case study protocol~\cite{Yin2018}. We preferred that the respondents use the user interface while the slides were a backup.  The order of themes differed as the interviews aimed at informal and open dialogue, but we followed the slides to keep track that all themes were covered.  We listened to the recordings later and  transcribed the key parts of the answers. For the transcribed answers, we carried out a narrative synthesis~\cite{cruzes2015case}, meaning that we used "words and text to condense and explain the findings of the synthesis". In practice, we collected and organized responses as qualitative  descriptions of the interview themes. We applied critical judgment in qualitatively selecting and synthesizing responses, such as how relevant a theme was for the respondent in their daily work, and how familiar a respondent was with the theme. We used the Microsoft OneNote notes-taking tool in the analysis as all data was text, the amount of data was moderate, and OneNote allows a free organization of themes and text in a two-dimensional space.  The results are summarized as follows.

\textbf{Results.}
The users appreciated very different functionalities, although they understood that the other functionalities could be important for other roles or tasks. For example, two users considered finding duplicates the key functionality while the others did not consider duplicate detection relevant to their daily work. 
The duplicate detection was also considered important for large projects and less so in small projects.
The existing dependencies and larger issue graphs are especially important and challenging for the R\&D team lead and product managers who value visualization.
A user summarized vividly: "\textit{Using Jira is like looking through a keyhole}". 

Although our solution relies on data projection from Jira that can be out of sync when issues are updated, the users commented that even day-old information is usable, although a practical update interval should be from a few minutes up to an hour, especially during the busy days before a release.

\textit{Issue graphs.}
The respondents liked the $p$-depth issue graph and its visualizations as a means of capturing information at a glance. 
The users considered depths 2--4 most relevant -- a 5-depth issue graph already showed them too much information.
One user discussed representing the parent-child hierarchy better while acknowledging that it is difficult to visualize without ending up with a very wide view and being a very implementation-specific challenge. 
Likewise, another user mentioned a release as another relevant viewpoint.
The users also commented on the user interface. A recurring comment concerned adding more information, such as tooltips or additional information, by hovering the mouse cursor. 

\emph{Dependency detection.}
Finding duplicated issues was considered the most practical technique although other types of missing dependencies were also acknowledged. The users felt that detection could take place in different phases and tasks, mentioning creating, triaging, resolving, and managing issues, and making releases. The time around releases is especially critical for finding duplicates, although the earlier the duplicates are found the better, especially if the reported issue turns out to be a blocker. Nobody considered false positive or incorrect proposals to be a problem because a proposal needs to be checked manually anyway, and proposals can always be disregarded -- false negatives or undetected proposals were considered much more inconvenient.
In particular, one user noted that duplicate detection could also be used to find similar older issues in order to find out how they were resolved or who resolved them so that users could be asked for help or even to resolve similar open issues. Our solution to store rejected dependency proposals and not show them again to any user was considered possible, although a more delicate approach could be applied. That is, a rejection decision is context- and sometimes user-specific, and it should be possible to revise the decisions. In particular, if an issue is changed, the rejection decision should be re-evaluated. Additional desired functionality was that the detectors should detect if issues have changed and the existing dependency between them has become obsolete. In contrast, predicting the type of dependency was not considered important or even feasible.

\emph{Consistency check.}
The users considered consistency checking to be relevant, especially in larger projects where the complexity and sizes of issue dependency networks have grown. Such a large project at TQC contains several parallel versions and multiple R\&D teams. In small projects, the users did not consider consistency checks necessary because the users can manage consistency manually. 
One user reported that, on the one hand, the consistency check would be more valuable if the processes inside TQC were more rigorous and issues contained fewer inconsistencies. On the other hand, he reckoned that the consistency check has the potential to improve the processes if inconsistencies or incorrect information can be made more visible. 
This could also make it possible to more reliably check cross-project dependencies.
A challenge for consistency check was said to be the time-boxed releases where the release is often set to the issues only after the resolving solution is ready -- if at all. 
Thus, for detected inconsistencies in issues, the corresponding resolving solutions need to be checked and might exist, meaning that a cause of the inconsistency is sometimes in the correspondence between Jira issues and their resolving solution. The limitation of the consistency check to the `parent-child', `requires', and `duplicate' dependencies was extensive enough. All respondents commented that only a general `relates' dependency would also be useful, but nothing additional was needed. Finally, other checks, such as identification of cyclic dependencies, could be interesting but not yet clearly needed in practice.

%\section{Results}
%\label{section:results}
%\input{results}

\section{Discussion}
\label{section:discussion}

\subsection{Discussion on RQ1: issue trackers main drawbacks}
\emph{RQ1. What  drawbacks  do  stakeholders  suffer  with current issue trackers?} 

When focusing on the constructs and the quality of the underlying issue dependency network, large, collaborative, long-lived projects bring forward the limitations of the issue trackers with respect to the data model (Drawback1), missing explicit dependencies  (Drawback2\&3), and inconsistencies  (Drawback4). This results in an incomplete broad view, critical for complex tasks like product management. The number of issues, potential dependencies, and stakeholders involved, all of them in constant change, raise the complexity. 

However, and as a consequence of this complexity, as our experiences from constructing and evaluating a solution to alleviate the drawbacks have taught us, capturing all dependencies and having complete consistency are elusive targets and even based on subjective and contextual judgment --- issues are not a static specification but a constantly evolving network of \emph{things to be done}. Thus, the drawbacks need to be mitigated rather than resolved. Therefore, it is vital to provide users with useful information and practical support features when using issue trackers, rather than aiming at fully automatic decision making. It is noteworthy that drawbacks are not necessarily TQC- or even Jira-specific but can be generalized to using other issue trackers, especially to those popular issue trackers with similar advanced features or other systems of similar use, appearing especially in the aforementioned large-project contexts.

\subsection{Discussion on RQ2: issue management features}
\label{sec:rq2-discuss}
\emph{RQ2. What features can be added to issue trackers to address these drawbacks?} \\

Our solution proposal of issue graphs forms a parallel, automatically constructed view of the data available in issue trackers, enabling more efficient dependency management and visualization (Objective1). Beyond the lifecycle of a single issue, we proposed to treat dependencies as first-class entities with their properties, which are usable, e.g., in dependency detection. We used issue ($r_0$) centered, bottom-up $p$-depth issue graphs ($G^p_0$) as the principal contextual structure for analysis, visualization, and users. The evaluation in the context of TQC's Jira showed the technical feasibility of issue graphs but also indicated that issue graphs up to depth five appear meaningful for users.  However, future work can allow other partial issue graphs and better emphasize existing hierarchies between issues. Interestingly, the issue tracker users did not consider the simple dependency typology of detectors a limitation but instead considered the existing typology too complex, opting for a typology of `duplicate', `parent-child', `requires', and possibly generic `relates' dependencies.

Regarding the extension techniques, the detection techniques  (Objective2) aim to assist users with simple but effective algorithms that operate with large data-sets. The evaluation showed that algorithms effectively propose dependencies and complement each other. Therefore, a quintessential system-view is  needed for the techniques and algorithms by contextualization that combines proposals, considers them in the context of existing issue graphs and issue properties, and manages rejected dependencies. This system view is based on the premise that a combination of many simple tools, such as voting and contextual adaptivity, is  comparable to a more advanced single algorithm, especially for domain experts. While this relatively simple but holistic solution appeared valuable, bringing forward many practical consequences, the solution can be further improved by more refined rejection handling and adding other --- more advanced --- detection techniques and algorithms, which can require a different aggregation approach. Another desired improvement is the explainability of detection techniques, pointing out why a proposal was made. While these findings are largely applicable to any issue tracker, the reference detection technique, as an example, relies on users' comments and quite Jira-specific textual IDs for which another complementary replacement technique may have to be constructed or adapted for other contexts. 

Regarding the consistency check and diagnoses (Objective3), evaluation of the techniques indicated that, rather than achieving complete consistency, these techniques' practical value is to make inconsistencies in an issue graph visible for the context a user is working on. This improves the transparency and control of the development process and can even induce process improvements. To this end, our consistency check and diagnoses techniques did not focus on fully-automated decision making but on providing users with assistance during the consistency check process within a specified ($G^{p=5}_0$) context of analysis rather than a complete analysis of all inconsistencies, which might not be relevant or even practical information.  The evaluation in the context of TQC's Jira also showed that the number of inconsistencies increases inconveniently rapidly for a user when the context of analysis grows. We argue that similar phenomena appear in other issue trackers of large projects due to the inherent characteristics of dependent and evolving issue tracker data.
Among the main future challenges are more suitable and efficient algorithms for diagnosis and a study of other analyses, such as redundant dependencies, including their practical value.

Thecess of applying the techniques beyond TQC relies somewhat on the characteristics of the project and Jira usage: the development and projects at TQC are relatively mature and Jira-centered, and the triage process assures a certain level of quality. As a result, TQC has a large amount of data in Jira, and Jira is actively used, including comments to reveal dependencies. 
TQC is also a medium-sized organization with hundreds of active Jira users, although there are open source and user communities that are less active. In other contexts, the techniques can be less successful, such as in the early phases of projects when there are fewer explicit requirements, the amount of requirements is small, the development practices are immature or not centered around Jira, or there is only a small group of stakeholders. Even at TQC, the stakeholders perceived the techniques differently, depending on their working context, such as being less valuable in small projects. Although techniques are scalable to large organizations of active users, the practical problems and value of solutions need to be assessed. However, an issue tracker is still mostly the repository of issues, the tool for issue workflow, and the tool for views to issues. That is, the issue trackers do not have much intelligence in their functionalities, and intelligent techniques can have some, although sometimes minor, benefits in any context.

\subsection{Discussion on RQ3: Artifact implementation}
\emph{RQ3. How can these features be integrated in an issue tracker in a way that it has value for use?} \\

The Jira plugin and microservice-based architecture we depicted in RQ3 address practical implementation of the techniques and use concerns. 
This plugin technology facilitates compatibility, security, and usability in the context of TQC's Jira.  However, TQC's Jira is standard deployment and, apart from the integration microservice (Milla), other microservices are independent of Jira, providing good maintainability, portability, and compatibility. The system should be deployable beyond TQC's Jira to other Jira installations and with minor modifications even  to other issue trackers and systems, such as requirements management, backlog, or roadmapping systems. We have already prototyped the same microservices in a research prototype. Likewise, we have prototyped two other, more advanced detectors within the system, which turned out to be too unreliable.

On the one hand, a solely plugin-based design could be done for a smaller data-set, but the design would have been very Jira-specific, resulting in an inefficient and more complex design. On the other hand, we had the microservices operational without plugin technology, but the microservices then could not handle the private issues, write decisions to Jira, or integrate the user interface with Jira. Such an independent tool from Jira was considered to have little practical value for TQC. 
The data projection was another key design decision that allowed us to separate batch processes and user queries. This was needed for the microservice-based solution and beneficial for efficiency, while the disadvantages were within users' acceptance limits. 

Besides the above improvements to the solution, certain design improvements could be considered. Our primary focus was not on graphical design and usability,  which can be improved. Additionally, the system's usability could be improved by integrating it into existing dashboards rather than operating as a separate plugin.

\subsection{Comparison to related work}

%Background \& related work summary available https://docs.google.com/spreadsheets/d/1W6N4h9shPgsW0gzJScWvYnT75DLRmo7ONfToAjVSo04/edit?usp=sharing

\textbf{Feature extension of traditional issue trackers in open source context.} Several studies have focused on analyzing the main challenges raised by traditional issue trackers in open source environments. Bertram et al.~\cite{Bertram2010} reported a list of seven design consideration features for issue trackers based on a qualitative study of their main drawbacks, including (i) providing customizable features for the visualization of issues data and their relations, and (ii) the simplification of tagging and reporting complex issue properties such as `requires' or `duplicates' relations, opening the door to automated features for the autonomous detection of these properties. 
Baysal et al.~\cite{Baysal2014} ran a qualitative analysis through 20 personal interviews with Bugzilla community stakeholders. The interviews identified that developers faced difficulties managing large issue repositories due to the constant flow of data (e.g., new issues, comments, reported dependencies) and the lack of support for filtering, visualizing, and managing changes in the issue dependency network. Heck and Zaidman~\cite{Heck2013} studied a set of 20 open source GitHub projects, from which they highlighted the management of duplicated issues and the visualization of the issues and issue dependencies as two of the most critical challenges for software developers. However, these contributions are limited to providing general highlights to key challenges and features for issue management tasks rather than designing and depicting concrete, detailed processes or theoretical models for the practical application of these features.

\textbf{Modeling and visualization of the issue dependency network.} Both Baysal et al. and the Heck and Zayman studies mentioned above highlight visualization of the issue dependency network beyond the single-issue perspective. %is highlighted by both Baysal et al. and Heck and Zaiman studies mentioned above. 
The latter narrowly depicts a modeling and visualization proposal based on the Bug Report Network (BRN) proposed by Sandusky et al.~\cite{Sandusky2004}, where an issue dependency network is represented as a tree of issues linked by their relations (including dependencies and duplicate relationships). The \textit{swarmOS Analyzer}\footnote{https://marketplace.atlassian.com/apps/1217806/} Jira plugin delivers a practical solution for representing the issue dependency network as an issue graph. Despite its filtering and classification features, it lacks advanced visualization tools to enable large projects to simplify and adapt the context of visualization to a specific issue or sub-set of issues.

\textbf{Dependency detection and duplicate detection in issue management.} Although the state-of-the-art addresses the requirements for traceability and dependency management, very few focus on the issue tracker domain. Borg et al.~\cite{Borg2014} conducted a systematic mapping of information retrieval techniques for traceability and artifact dependencies in software projects. Among 79 related publications, most were limited to a proof-of-concept solution with a reduced sample validation with partial quality metrics like precision or recall in a validation scenario of no more than 500 artifacts. Despite the supporting tools like Jira plugins for the visualization of issue dependency trees, like SwarmOS Analyzer or Vivid Trace\footnote{https://marketplace.atlassian.com/apps/1212548/}, there are no popular plugins or tools for the autonomous detection of dependencies or cross-references among issues in an issue repository. 

On the other hand, managing and detecting duplicated issues is a well-known problem considered critical by several studies when managing issues with issue trackers~\cite{Alipour2013,Kshirsagar2015,deshmukh2017}. Ellmann~\cite{Ellmann2018} defines a theoretical background for the potential of state-of-the-art natural language and machine learning techniques to extend issue trackers with automated duplicate detection. However, no artifact nor practical implementation is reported. The \textit{Find Duplicates}\footnote{https://marketplace.atlassian.com/apps/1212706/} Jira plugin uses similar techniques to those reported by Ellmann to extend search features from Jira by reporting potential duplicates at report time or running queries to find related issues. Nevertheless, these tools do not provide details about the scalability of these solutions for large data-sets, as the emphasis is on proof-of-concept evaluation. Instead, they offer centralized server-side extensions for Jira environments with few details from a software architecture point of view, making them less suitable for large data-sets.

\textbf{Consistency check and repair of releases.} As reported in Section~\ref{sec:drawbacksManagement-techniques}, literature on release planning for issue management is especially focused on autonomous release plan generation rather than consistency checking and repair of releases~\cite{Svahnberg2010,Ameller2016}. Consequently, it is difficult to find related work focused on analyzing and diagnosing releases in the issue tracker domain. If we focus on tool support examples, in addition to the visualization of issue dependencies, the~\textit{Vivid Trace} Jira plugin uses this feature to provide deep dependency analysis capabilities focused on visual representation, monitoring of chains of events, and the detection of potential blockers or conflicts among the dependencies.

\section{Threats to validity}
\label{sec:Validity}
We analyze the threats to validity according to the four categories proposed by~\cite{shadish2002experimental} in experimental research.

\emph{Construct validity} refers to proper conceptualization or theoretical generalizations. This study focused on tool (Jira) improvement rather than process improvements. Our conceptualization is based on a few stakeholders, and, as noted in the validation interviews, their needs differ. One threat is whether we conceptualized the problem correctly, and another is whether we focused on a relevant problem of the case company.  However, the respondents were highly experienced, there were several of them, the researcher had a prolonged engagement with the problems as the process lasted a reasonably long time, and the problems the experts raised were also evident in the data. Furthermore, the results cause no harm either as they aim to help and do not disturb existing ways of working.  
In our solution development, we relied on handpicked examples. In order to alleviate potential threats with the selection of the examples, we established good communication with TQC’s stakeholders. In eliciting the drawbacks in RQ1, we used carefully designed and piloted interviews. This helped us assess which issues would be suitable examples for our research. However, the evaluation iterated through all public data, except for cross-validation, thus not limiting ourselves to the hand-chosen examples.

\emph{Internal validity} refers to inferences about whether the presumed treatment and the presumed outcome reflect a causal relationship between them. Our solution aims to address drawbacks acknowledged beforehand by the stakeholders. Thus, the knowledge claim concerns whether the suggested solution, i.e., techniques implemented and integrated into Jira, helps address the drawbacks. The solutions were validated with TQC’s Jira users to check that they were applicable to tackle the drawbacks. However, a limitation is that the Jira users testing our system used real data but did not test it extensively in their daily work.

\emph{External validity} concerns whether our knowledge claims could be generalized beyond the TQC environment. We consider TQC a good case for research due to its large, standard Jira and typical software engineering and open source practices. Thus, there is a high probability that the solutions could be applicable in other environments. However, TQC’s Jira is a mature and complex environment, and the drawbacks and our solutions reflect this. Although our solutions may technically work in less complex environments, it is not certain that they would be equally valuable. In terms of the mutability of the artifact, we intentionally constructed the solution to be flexibly adaptable to new algorithms and microservices. Interviews with a few selected users do not fully compare to full-scale use in practice. This is notable as the generalizability of the artifact is, in addition to its applicability to the drawbacks themselves, also dependent on whether the users accept the solution. This is difficult to assess with only a few respondents and might come down to, for example, whether or not the users are satisfied with the artifact and its microservices in the long run, and not just initially. 

\section{Conclusions and Future Work}
\label{section:conclusions}
We have presented an approach that addresses drawbacks in an issue dependency network.
The contributions are in applied Design Science research in the context of the use of issue trackers in large projects that TQC's Jira concretizes. 
The basis of the solution is having issues and dependencies as separate objects and automatically constructing a complimentary issue graph. 
The dependency detection complements the issue graph by proposing missing dependencies, and the consistency check identifies incorrectness in the issue graph.
The results show how to adopt the relatively technologically straightforward techniques in a complex collaborative issue tracker use-context and a large data-set, considering the integrated system concern, practical applicability, and inherent incompleteness of issue data. 
%The solution aims to mitigate the drawbacks rather than resolve them.
%
The system is not yet in active use because it is a research prototype without guaranteed technical support and maintenance for TQC. However, TQC has expressed interest in having the system operational, and the results can be generalized beyond TQC.

This research was carried out in the context of TQC's Jira, and we can address several directions for future work.  
Issue trackers still appear to be a little-researched area, although they are prevalent in open source communities and widely used in other organizations. 
Qt is a dual licensed open source project governed by TQC, although not all maintainers work at TQC. The commercial license provides better support and code under a different (non-GPL) license rather than necessarily more features except for specific industries in the private projects.
According to our understanding, many open source projects are developed similarly in the sense that there is typically a core group of maintainers and developers who might even work in some big companies. 
Therefore, our results seem generalizable to mature open source projects and medium-sized companies, which have up to a few hundred active developers and maintainers. However, the generalization would benefit further assessment. The applicability and benefits of the techniques need further studies in small or very large organizations and more immature or early-phase development projects.  More research on issue trackers can be similarly conducted, including studies on how they are used and adding intelligence to their functionalities that would benefit different development practices and organizations.

\bibliographystyle{IEEEtran}
\bibliography{mainTSEOpenReqQT}

  \section*{Acknowledgments}

The work presented in this paper has been conducted within the scope of the Horizon 2020 project OpenReq, which is supported by the European Union under Grant Nr. 732463. We are grateful for the provision of the Finnish computing infrastructure to carry out the tests  (persistent identifier urn:nbn:fi:research-infras-2016072533). This paper has been funded by the Spanish Ministerio de Ciencia e Innovación under project / funding scheme PID2020-117191RB-I00 / AEI/10.13039/501100011033.

\vskip 4\baselineskip plus -1fil 

\begin{IEEEbiographynophoto}{Mikko Raatikainen} received his PhD in computer science and engineering from Aalto University. He is a researcher of the empirical software engineering research group in  University of Helsinki. His research interests include empirical research in software engineering and business.
\end{IEEEbiographynophoto}

\vskip -2\baselineskip plus -1fil 

\begin{IEEEbiographynophoto}{Quim Motger} is a PhD student at Universitat Politècnica de Catalunya (UPC). He is a member of the UPC research group on software and service engineering. His research focuses on natural language processing, machine/deep learning software systems, and web-based software architecture environments.
\end{IEEEbiographynophoto}

\vskip -2\baselineskip plus -1fil 

\begin{IEEEbiographynophoto}{Clara Marie Lüders} is a PhD student at University of Hamburg (UHH). She is a member of the UHH research group on applied software technology. Her research focuses on machine/deep learning, natural language processing, Issue Tracking Systems, and graph theory.
\end{IEEEbiographynophoto}

\vskip -2\baselineskip plus -1fil 

\begin{IEEEbiographynophoto}{Xavier Franch} received his PhD from Universitat Politècnica de Catalunya (UPC). He is a full professor in UPC where he leads the research group on software and service engineering. His research focuses on requirements engineering and empirical software engineering. He is associate editor in IST, REJ, and Computing, and J1 chair at JSS.
\end{IEEEbiographynophoto}

\vskip -2\baselineskip plus -1fil

\begin{IEEEbiographynophoto}{Lalli Myllyaho} is a PhD student at University of Helsinki (UH). With a background in mathematics and teaching, he is a member of the empirical software engineering group at UH. His current interests include the reliability and operations of machine learning systems.
\end{IEEEbiographynophoto}

\vskip -2\baselineskip plus -1fil 

\begin{IEEEbiographynophoto}{Elina Kettunen} received her PhD in plant biology and her Master's degree in computer science from University of Helsinki. Her research interests include empirical software engineering and paleobotany.
\end{IEEEbiographynophoto}

\vskip -2\baselineskip plus -1fil

\begin{IEEEbiographynophoto}{Jordi Marco} received his Ph.D. from Universitat Politècnica de Catalunya (UPC). He is an Associate Professor in Computer Science at the UPC and a member of the software and service engineering group (GESSI). His research interests include natural language processing, machine learning, service-oriented computing, quality of service, and conceptual modeling. 
\end{IEEEbiographynophoto}

\vskip -2\baselineskip plus -1fil 

\begin{IEEEbiographynophoto}{Juha Tiihonen} received his PhD in computer science and engineering from Aalto University. His research interests include configuration systems and processes for physical, service, and software products. This work was performed at University of Helsinki. 
He is currently the lead developer of sales configuration systems at Variantum~oy.
\end{IEEEbiographynophoto}

\vskip -2\baselineskip plus -1fil 

\begin{IEEEbiographynophoto}{Mikko Halonen} is a B.Sc (Automation Eng. Tech.) from the Technical College of Oulu. He currently works as a quality manager in The Qt Company.
\end{IEEEbiographynophoto}

\vskip -2\baselineskip plus -1fil 

\begin{IEEEbiographynophoto}{Tomi Männistö} received his PhD from Helsinki University of Technology, currently Aalto University. He is a full professor of the empirical software engineering research group in University of Helsinki. His research interests include software architectures, variability modelling and management, configuration knowledge, and requirements engineering.
\end{IEEEbiographynophoto}

\end{document}